\documentclass{edm_article}

 \usepackage{url}
\usepackage{fixmath}
\usepackage{amsmath}
\usepackage{mathrsfs}
\usepackage{algorithm2e}
\usepackage{hyperref}
\usepackage{bbm}
\usepackage{graphicx}
\usepackage{subfigure}
\usepackage{adjustbox}
\usepackage{multirow}
\usepackage{stfloats}
\usepackage{xcolor}
\begin{document}

\title{Modeling Knowledge Acquisition from Multiple Learning Resource Types}

%\subtitle{[Extended Abstract]
%\titlenote{A full version of this paper is available as
%\textit{Author's Guide to Preparing ACM SIG Proceedings Using
%\LaTeX$2_\epsilon$\ and BibTeX} at
%\texttt{www.acm.org/eaddress.htm}}}
%
% Submissions for EDM are double-blind: please do not include any
% author names or affiliations in the submission. 
% Anonymous authors:

\numberofauthors{3}
\author{
\alignauthor
Siqian Zhao\thanks{First two authors contributed equally to this work.}\\
     \affaddr{Computer Science}\\
    %   \affaddr{Department of Computer Science}\\
      \affaddr{University at Albany, SUNY}\\
      \affaddr{Albany, NY 12222 USA}\\
      \email{szhao2@albany.edu}
\alignauthor
Chunpai Wang\footnotemark[1]\\
     \affaddr{Computer Science}\\
    %   \affaddr{Department of Computer Science}\\
      \affaddr{University at Albany, SUNY}\\
      \affaddr{Albany, NY  USA}\\
      \email{cwang25@albany.edu}
\alignauthor
Shaghayegh Sahebi\\
     \affaddr{Computer Science}\\
    %   \affaddr{Department of Computer Science}\\
      \affaddr{University at Albany, SUNY}\\
      \affaddr{Albany, NY 12222 USA}\\
      \email{ssahebi@albany.edu}
}

\maketitle

%\onecolumn

\begin{abstract}
% -The recent growth of online learning environments and Massive Open Online Courses (MOOCs) has led to diversity and variety of learning materials, such as video lectures, multiple-choice questions, reading materials, and discussions.
Students acquire knowledge as they interact with a variety of learning materials, such as video lectures, problems, and discussions. Modeling student knowledge at each point during their learning period and understanding the contribution of each learning material to student knowledge are essential for detecting students' knowledge gaps and recommending learning materials to them. Current student knowledge modeling techniques mostly rely on one type of learning material, mainly problems, to model student knowledge growth. These approaches ignore the fact that students also learn from other types of material.
% -, and hence lose useful information in modeling student knowledge.
In this paper, we propose a student knowledge model that can capture knowledge growth as a result of learning from a diverse set of learning resource types while unveiling the association between the learning materials of different types. Our multi-view knowledge model (MVKM) incorporates a flexible knowledge increase objective on top of a multi-view tensor factorization to capture occasional forgetting while representing student knowledge and learning material concepts in a lower-dimensional latent space. We evaluate our model in different experiments to %by predicting students' performance, examining learned student knowledge growth, and analyzing learning materials' similarities. 
% \textcolor{red}{discovering student latent similarities}, 
%Our experiments 
show that it can accurately predict students' future performance, differentiate between knowledge gain in different student groups and concepts, and unveil hidden similarities across learning materials of different types. 
\end{abstract}

%% A category with the (minimum) three required fields
%\category{H.4}{Information Systems Applications}{Miscellaneous}
%%A category including the fourth, optional field follows...
%\category{D.2.8}{Software Engineering}{Metrics}[complexity measures, performance measures]
%
%\terms{Theory}

 % NOT required for Proceedings

\maketitle

\section{Introduction}
\label{sec:intro}
Both student knowledge modeling and domain knowledge modeling are important problems in the educational data mining community.
%Modeling student knowledge helps in predicting student performance in the next quizzes, identifying students that are at risk of dropout, determining student knowledge gaps, and recommending appropriate learning resources to students.
In this context, student knowledge tracing and knowledge modeling approaches aim to evaluate students' state of knowledge or quantify students' knowledge in the concepts that are presented in learning materials at each point of the learning period~\cite{Corbett1994,Baker2008,Yudelson2013,Khajah2014,Zhang2017a,Nagatani2019,Choffin2019,Vie2019}.
% answer questions such as what knowledge is offered in a learning resource or required to solve a problem (domain knowledge modeling), what is a student’s state of knowledge at each point of learning period (student knowledge modeling), and how a specific student would perform on a test (predicting student performance).
Domain knowledge modeling, on the other hand, focuses on understanding and quantifying the topics, knowledge components, or concepts that are presented in the learning material~\cite{barnes2005q, casalino2017q, lan2014time}.
It is useful in creating a coherent study plan for students, modeling students' knowledge, and analyzing students' knowledge gaps.
% In the presence of abundant online learning materials and high cost of labeling by human experts, automatic domain knowledge models are essential for the feasibility of educational data mining solutions.

%% ------------------------- a successful knowledge modeling should be:  ------------------------- 
% personalized
% knowledge gain/forgetting, that is associated with student activities and our observation of their success
A successful student knowledge model should be personalized to capture individual differences in learning~\cite{Yudelson2013,Lan2014c}, understand the association and relevance between learning from various concepts~\cite{Sahebi2016b,Zhang2017a}, model knowledge gain as a gradual process resulting from student interactions with learning material~\cite{Gonzalez-Brenes2013a,Piech2015,Doan2019a}, and allow for occasional forgetting of concepts in students~\cite{Choffin2019,Nagatani2019,Doan2019a}.
%% ------------------------- but current approaches have two main issues  ------------------------- 
Despite recent success in capturing these complexities in student knowledge modeling, a simple, but important aspect of student learning is still under-investigated: that students learn from different types of learning materials.
%the current student knowledge modeling approaches suffer from one or both of the following issues: modeling a single learning resource type, and requiring a predefined domain knowledge model.
% modeling just one source of knowledge
%First, the 
Current research has focused on modeling one single type of learning resource at a time (typically, ``problems''), ignoring the heterogeneity of learning resources from which students may learn. 
Modern online learning systems frequently offer students to learn and assess their knowledge using various learning resource types, such as readings, video lectures, assignments, quizzes, and discussions. 
Previous research has demonstrated considerable benefits of interacting with multiple types of materials on student learning. 
For example, worked examples can lead to faster and more effective learning compared to unsupported problem solving~\cite{Najar2014}; and enriching textbooks with additional forms of content, such as images and videos, increases the helpfulness of learning material~\cite{Agrawal2011,Agrawal2014}. 
Ignoring diverse types of learning materials in student knowledge modeling limits our understanding of how students learn.

One of the obstacles in considering the combined effect of learning material types is the lack of explicit learning feedback from all of them. Some learning material types, such as problems and quizzes, are gradable.
As students interact with such material types, the system can perceive student grade as an explicit feedback or indication of student knowledge: if a student receives a high grade in a problem, it is likely that the student has gained enough knowledge required to solve that problem. 
On the other hand, some of the learning materials are not gradable and their impact on student knowledge cannot be explicitly measured.
For example, we cannot directly measure the consequent knowledge gain from watching a video lecture or studying an example. 

As an alternative for quantifying student knowledge gain, the system can measure other quantities, such as the binary indication of student activity with a learning material or the time they spent on it. 
However, this kind of measure may result in contradictory conclusions~\cite{Beck2008,Huang2015a,Hosseini2016}. 
For example, spending more time to study the examples provided by the system may both increase the student's knowledge, and at the same time, be an indicator of a weaker student, who does not have enough knowledge in the provided concepts.
These weaker students may select to study more examples to compensate for their lower knowledge levels.
Consequently, the knowledge gain of studying these auxiliary learning materials is usually overpowered by the student selection bias and is not represented correctly in the overall dataset.

A similar issue exists in the current domain knowledge models. 
The automatic domain knowledge models that are based on students' activities mainly model one type of learning material and ignore the relationship between various kinds of learning materials~\cite{doan2019rank, casalino2017q}.
Alternatively, an ideal domain knowledge model should be able to model and discover the similarities between learning materials of different types. %

%Ideally, we should understand the connections among various learning material types and the knowledge span offered by combinations of
%learning material to guide students effectively towards the right learning path. 
%For example, if a student has trouble solving a problem, we should be able to suggest this student to watch a related video or work on an associated example.

% need domain knowledge model

%% ------------------------- we resolve those issues with a simple approach  ------------------------- 
% personalized by having different student vectors
% allow/learn forgetting 
% learn domain model while knowledge gain
% model learning from multiple knowledge resources

%Student interaction with these different types of learning materials can contribute differently to student knowledge. 
In this paper, we simultaneously address the problems of student knowledge modeling and domain knowledge modeling, while considering the heterogeneity of learning material types.
We introduce a new student knowledge model that is the first to concurrently represent student interactions with both graded and non-graded learning material.
Meanwhile, we discover the hidden concepts and similarities between different types of learning materials, as in a domain knowledge model.
To do this, we pose this concurrent modeling as a multi-view tensor factorization problem, using one tensor for modeling student interactions with each learning material type. 
By experimenting on both synthetic and real-world datasets, we show that we can improve student performance prediction in graded learning materials, as measured by the Root Mean Squared Error (RMSE) and Mean Absolute Error (MAE).  %TODO 
In summary, the contributions of this paper are:\\
% \begin{itemize}
    1) proposing a personalized, multi-view student knowledge model (MVKM) that can capture learning from \textit{multiple learning material types} and allow for occasional student forgetting, while modeling all types of learning materials;\\
    2) conducting experiments on both synthetic and real-world datasets showing that our proposed model outperforms conventional methods in predicting student performance;\\
    3) examining the resulting learning material and student knowledge latent features to show the captured similarity between learning material types and interpretability of student knowledge model.
% \end{itemize}

\section{Related Work}
\label{sec:related}
% -\noindent\textbf{Student Performance Prediction and Knowledge Modeling}
\noindent\textbf{Knowledge Modeling}
% -In student performance prediction, the goal is to estimate a student's grade, score, success, or failure in graded learning materials, such as problems and quizzes~\cite{Desmarais2012}. 
% -Although this problem can be addressed without explicitly modeling students' knowledge state~\cite{PavlikJr2009}, having an understanding of student knowledge can be useful in comprehending the fundamental reasons of student success or failure. 
Student knowledge modeling aims to quantify student knowledge state in the concepts or skills that are covered by learning materials at each learning point.

Pioneer approaches of student knowledge modeling, despite being successful, were not personalized, relied on a predefined (sometimes expert-labeled) set of concepts in learning material, did not allow for learned concepts to be forgotten by students, and modeled each concept independently from one another~\cite{Drasgow1990,Corbett1994,Pavlik2009,Linden2013}.
%In other words, in these models all students were assumed to have similar progress according to their interaction with learning materials.
%Also, assuming that a correct mapping between learning material and concepts exists, these models left the burden on experts to manually label the material~\cite{Pardos2010a,Wilson2016}, or models that are solely for 
Later, some student knowledge models aimed to solve these shortcomings by learning different parameters for each (type of) student~\cite{Pardos2010a,Yudelson2013,Koedinger2013a}, including decays to capture forgetting of concepts in learner models~\cite{Qiu2011,Lindsey2014a,Mozer2016} and capturing the relationship between concepts that are present in a course~\cite{Thai-Nghe2011,Gonzalez-Brenes2013a}.
Yet, these models assume that a correct domain knowledge model, that maps learning material into course concepts, exists.

In recent years, new approaches aim to learn both domain knowledge model and student knowledge model at the same time~\cite{Lan2014c,Gonzalez-Brenes2015,Sahebi2016b,Wang2017,Zhang2017a,Doan2019a}.
Our proposed model falls into this latest category as it does not require any manual labeling of learning materials, while having
the ability to use such information if they are available.
It is personalized by learning lower-dimensional student representations, allows forgetting of concepts during student learning by adding a rank-based constraint on student knowledge, and models the relationship between learning material.
%-In addition to the above, our approach understands the relation between learning materials of different types and models student learning from multiple learning material types at the same time.

%\subsection{Domain Knowledge Modeling}

\noindent\textbf{Learning from Multiple Material Types}
In the educational data mining (EDM) literature, learning materials are provided in various types, such as problems, examples, videos, and readings. 
While there have been some studies in the literature on the value of having various types of learning materials for educating students~\cite{Agrawal2011,Beck2008,Najar2014}, the relationship between these material types, and their combined effect on student knowledge and student performance is under-investigated.

Multiple learning material types have been studied in the literature in finding insights into different activity distributions or cluster patterns between high-grade and low-grade students~\cite{Velasquez2014,Wen2014}, have been used as contextual features in scaffolding or choosing among the existing student models~\cite{Yudelson2008,SaoPedro2013}, have been added to improve existing domain knowledge models only for graded material types while ignoring student sequences~\cite{Botelho2016,Chen2016,Desmarais2015,Liu2016,Pardos2013,Sahebi2018,Pelanek19}, or have been classified into beneficial or non-beneficial for students~\cite{Alexandron2015}.
However, to the best of our knowledge, none of these studies have explicitly modeled the contribution of various kinds of learning materials on student knowledge during the learning period, the interrelations among these learning materials, and their effect on student performance.
% -One reason for unpopularity of using heterogeneous material types for predicting student performance is their potential conflicting effects. 
% -For example, Beck et al. investigated if providing assistance (help) to students benefits them using experimental trials, Bayesian Evaluation and Assessment framework, and learning decomposition~\cite{Beck2008}. 
% -In their studies, experimental trials and learning decomposition showed that assistance hurts students' learning. However, 
 The Bayesian Evaluation and Assessment framework found that assistance promoted students' long-term learning. 
More recently, Huang et al. discovered that adaptation of their framework (FAST) for student modeling by including various activity types may lead researchers to contradictory conclusions~\cite{Huang2015a}. 
More specifically, in one of their formulations student example activity suggests a positive association with model parameters, such as probability of learning, while in another formulation this type of activity has a negative association with model parameters. 
Also, Hosseini et al. concluded that annotated examples show a negative relationship with students' learning, because of a selection effect: while annotated examples may help students to learn, weaker students may study more annotated examples~\cite{Hosseini2016}.
%Another complication for considering heterogeneous material types is the difficulty in interpretation of students' observed activities, as the feedback that system receives. In graded learning material types, such
%as assignments and quizzes, a student's score explicitly represents her knowledge on the topic. Whereas in
%other material types, such as reading materials, there is no direct evaluation and no explicit observation of
%student's knowledge. Hence, measuring the effect of such learning materials on students' knowledge, and
%thus predicting their future performance, remains a challenging task.
The model proposed in this paper considers student interactions from multiple learning material types, mitigating over-estimation of student knowledge by transferring information from interactions with graded material, while accounting for knowledge increase that happen as a result of student interaction with non-graded material.

\section{Multi-View Knowledge Modeling}
\vspace{-1pt}
\label{sec:mvrbtf}
% share code + synthetic data
\subsection{Problem Formulation and Assumptions}
\label{sec:formulation}
We consider an online learning system in which $M$ students interact with and learn from multiple types ($r \in \mathfrak{R}$) of learning materials.
Each learning material type $r$ includes a set of $P^{[r]}$ learning materials.
A material type can be either graded or non-graded. 
Students' normalized grade in tests, success or failure in compiling a piece of code, or scores in solving problems are all examples of graded learning feedback.
Whereas, watching videos, posting comments in discussion forums, or interacting with annotated examples are instances of non-graded learning feedback that the system can receive.
We model the learning period as a series of student attempts on learning materials, or time points ($a \in \mathcal{A}$).
To represent student interaction feedback with learning materials of each type $r$ during the whole learning period $\mathcal{A}$, we use a $M\times P^{[r]}\times A$ three-dimensional tensor $\mathbold{X}^{[r]}$.
The $a^{\text{th}}$ slice of tensor $\mathbold{X}^{[r]}$, denoted by ${X}_a^{[r]}$, is a matrix representing student interactions with the learning material type $r$ during one snapshot of the learning period.
The $s^{\text{th}}$ row of this interaction matrix $\mathbold{x}_{a,s}^{[r]}$ shows feedback from student $s$'s interactions with all learning materials of type $r$ at attempt $a$; and the tensor element $x_{a,s,p}^{[r]}$ is the feedback value of student $s$'s activity on learning material $p$ of type $r$ at learning point $a$.

%We assume that students freely choose to interact with different learning materials and do not necessarily follow the same learning path.
%We assume that 
We use the following assumptions in our model:
\textbf{(a)} Each learning material covers some concepts that are presented in a course;
the set of all course concepts are shared across learning materials; and
the training data does not include the learning materials' contents nor their concepts.% represented by Q
\textbf{(b)} Different learning materials have different difficulty or helpfulness levels for students.
For example, one quiz can be more difficult than another one, and one video lecture can be more helpful than the other one. % represented by bias_q
% This can result in a general trend over student scores during the whole course.
\textbf{(c)} The course may follow a trend in presenting the learning material: going from easier concepts to more difficult ones or alternating between easy and difficult concepts; % represented by bias_a
despite that, students can freely interact with the learning materials and are not bound to a specific sequence.
\textbf{(d)} As students interact with these materials, they learn the concepts that are presented in them; meaning that their knowledge in these concepts increases. % represented by rank constraint on T
\textbf{(e)} Since students may forget some course concepts, this knowledge increase is not strict. % represented by rank constraint on T
\textbf{(f)} Different students come with different learning abilities and initial knowledge values. % represented by bias_s
\textbf{(g)} The gradual change of knowledge varies among different students. But, students can be grouped together according to how their knowledge changes in different concepts, e.g., some students are fast learners compared to others.  % represented by S
\textbf{(h)} Eventually, a student's performance in a graded learning material, represented by a score, depends on the concepts covered in that material, student's knowledge in those concepts, the learning material difficulty/helpfulness, and the general student ability. % represented by STQ + ...

In addition to the above, we have an essential assumption \textbf{(i)} that connects the different parts of our model: a student's knowledge that is obtained from interacting with one learning material type is transferable to be used in other types of learning materials.
In other words, students' knowledge can be modeled and quantified in the same latent space for all different learning material types.  
In the following, we first propose a single-view model for capturing the knowledge gained using one type of learning material (MVKM-Base) and then extend it to a multi-view model that can represent multiple types of learning materials.
%the knowledge gain, which is obtained by each learning material type, can be transferred to be used 

\subsection{MVKM Factorization Model}
% The model we are proposing is a simple extension of Rank-Based Tensor Factorization (RBTF)~\cite{Doan2019a} to multiple types of learning materials. 
% To add to the interpretability of our model, we also add the non-negativity and normalization constraints on the learning material embeddings. 
% RBTF has been shown to be successful in predicting student performance interacting with only graded learning materials, but has not been evaluated on student knowledge modeling nor in figuring out learning material latent concepts.
\noindent\textbf{The Proposed Base Model (MVKM-Base).} Following the mentioned assumptions in Section~\ref{sec:formulation}, particularly assumptions \textbf{(a)}, \textbf{(g)}, and \textbf{(h)}, and assuming that students interact with only one learning material type, %and inspired by use of collaborative filtering approaches in predicting student performance~\cite{Koren2009,Thai-Nghe2015,Doan2019a}, 
we model student interaction tensor $\mathbold{X}$ as a factorization ($n$-mode tensor product) of three lower-dimensional representations: 1) an $M\times K$ student latent feature matrix $S$, 2) a $K\times C\times A$ temporal dynamic knowledge tensor $\mathbold{T}$, and 3) a $C \times P$ matrix $Q$ serving as a mapping between learning materials and course concepts. 
In other words, we have $\hat{x}_{s,a,p} \approx \mathbold{s}_s\cdot T_{a} \cdot \mathbold{q}_p$.
Matrix $S$ here represents students being mapped to latent learning features that can be used to group the students (assumption \textbf{(g)}).
Tensor $\mathbold{T}$ quantifies the knowledge growth of students with each learning feature in each of the concepts while attempting the learning material.
Accordingly, the resulting tensor from product $\mathbold{K} = S\mathbold{T}$ represents each student's knowledge in each concept at each attempt.

\begin{figure}[]
    \centering
    \includegraphics[scale=0.45]{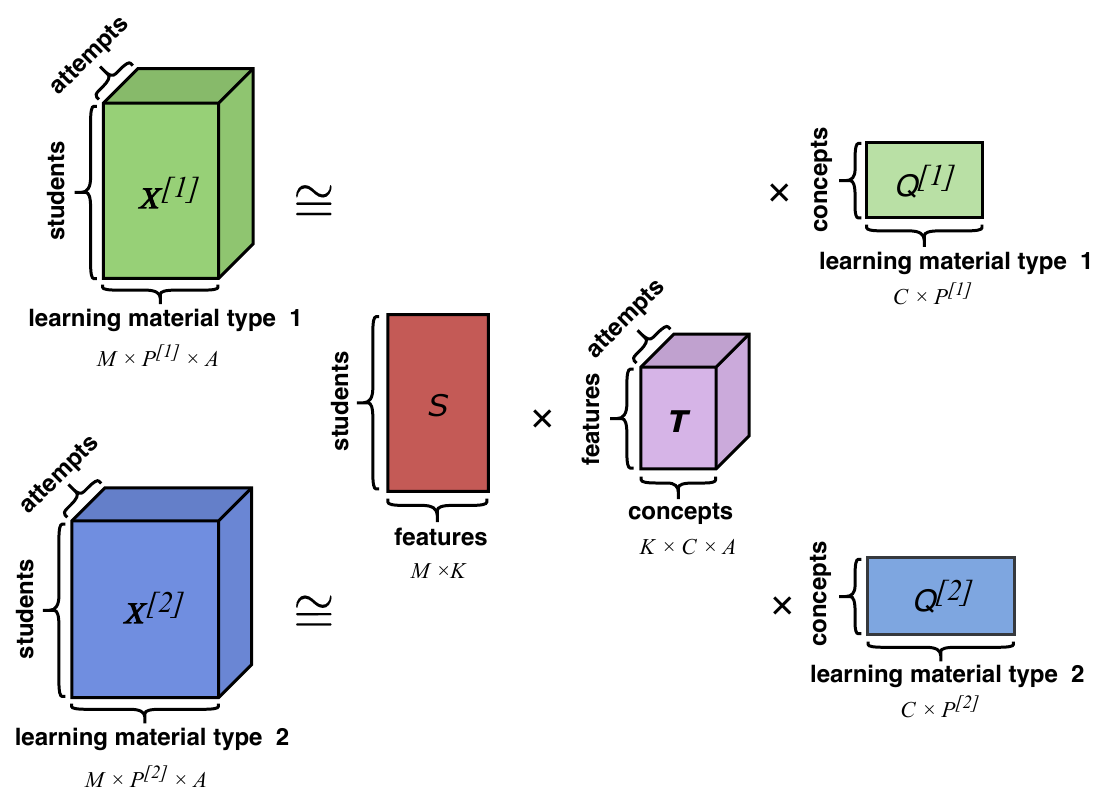}
    % \caption{Decomposing student interaction tensors with quizzes ($\mathbold{X}^{[1]}$) and discussions ($\mathbold{X}^{[2]}$) into a shared student knowledge tensor ($\mathbold{T}$) and separate quiz ($Q^{[1]}$) and discussion ($Q^{[2]}$) concept mappings}
    \vspace{-20pt}
     \caption{Decomposing student interaction tensors with two learning material types $\mathbold{X}^{[1]}$ and $\mathbold{X}^{[2]}$.}
    %  into shared latent student feature matrix $S$, dynamic knowledge tensor $\mathbold{T}$ and separate learning material concept mappings $Q^{[1]}$ and $Q^{[2]}$.}
    %  Suppose we have $M$ students, $P^{[1]}$ learning materials type 1, $P^{[1]}$ learning materials type 2, $A$ attempts, $K$ latent students' features and $C$ latent learning materials' features.}
    \label{fig:model}
\vspace{-10pt}    
\end{figure}

\vspace{-2pt}

To increase interpretability, we enforce the contribution of different concepts in each learning material to be non-negative and sum to one.  
Similarly, we enforce the same constraints on each student's membership in the student latent features.
Since each student can have a different ability (assumption \textbf{(f)}) and each learning material can have its own difficulty level (assumption \textbf{(b)}), we add two bias terms to our model ($b_s$ for each student $s$, and $b_p$ for each learning material $p$) to account for such differences.
To capture the general score trends in the course (assumption \textbf{(c)}), we add a parameter $b_a$ for each attempt. 
% $\textcolor{red}{\mu}$..
%Additionally, \textcolor{red}{as the inherent difficulties on different questions could be different we add a parameter $b_q$ for each question. }
Accordingly, we estimate student $s$'s score in a graded learning material $p$ at attempt $a$ ($\hat{x}_{a,s,p}$) as in Equation~\ref{eq:simplemodelbias1}.
% \textcolor{blue}{
% Here, $\mathbold{t}_{a,s}$ represents student $s$'s knowledge vector at attempt $a$, $\mathbold{q}^{[r]}_p$ shows material $p$'s concept vector, and $\mu^{[r]}$ captures the cohort bias or global average over all scores.}
Here, $T_{a}$ is a matrix capturing the relationship between student features and concepts at attempt $a$, 
$\mathbold{s}_s$ represents student $s$'s latent feature vector, 
$\mathbold{q}_p$ shows material $p$'s concept vector.
\vspace{-1pt}
\begin{equation}
% \hat{x}_{s,a,p} \approx \mathbold{s}_{s}\cdot \mathbold{T}_{a} \cdot \mathbold{q}_p + b_s + b_p + b_a + \mu^{[r]}
\hat{x}_{s,a,p} \approx \mathbold{s}_s\cdot T_{a} \cdot \mathbold{q}_p + b_s + b_p + b_a
\label{eq:simplemodelbias1}
\end{equation}
\vspace{-20pt}
% \begin{equation}
% % \hat{x}^{[r]}_{s,a,p} \approx \mathbold{s}_{s}\cdot \mathbold{T}_{a} \cdot \mathbold{q}^{[r]}_p + b_s + b^{[r]}_p + b^{[r]}_a + \mu^{[r]}
% \hat{x}^{[r]}_{s,a,p} \approx \mathbold{s}_s\cdot T_{a} \cdot \mathbold{q}^{[r]}_p + b_s + b^{[r]}_p + b_a
% \label{eq:simplemodelbias1}
% \end{equation}

We use a sigmoid function $\sigma(\cdot)$ to estimate student interaction with a non-graded learning material, or graded ones with binary feedback: 
$$
\hat{x}_{s,a,p} \approx \sigma (\mathbold{s}_s \cdot \mathbold{T}_{a} \cdot \mathbold{q}_p + b_s + b_p + b_a + \mu)
\label{eq:simplemodelbias2}
$$
% \vspace{-1pt}
% \begin{align*}
% \hat{x}_{s,a,p} \approx \sigma (\mathbold{s}_s \cdot T_{a} \cdot \mathbold{q}_p + b_s + b_p + b_a)
% \label{eq:simplemodelbias2}
% \end{align*}

% $$
% \hat{x}^{[r]}_{s,a,p} \approx \sigma (\mathbold{s}_s \cdot T_{a} \cdot \mathbold{q}^{[r]}_p + b_s + b^{[r]}_p + b_a)
% \label{eq:simplemodelbias2}
% $$
\vspace{-10pt}
\noindent\textbf{Modeling Knowledge Gain while Allowing Forgetting.} So far, this simple model captures latent feature vectors of students and learning materials, and learns $T$ as a representation of knowledge in students.
However, it does not explicitly model students' gradual knowledge gain (assumption \textbf{(d)}).
We note that students' knowledge increase is associated with the strength of concepts in the learning material that they interact with.
As students interact with learning materials with some specific concepts, it is more likely for their predicted scores in the relevant learning materials to increase.
With a Markovian assumption, we can say that if students have practiced some concepts, we expect their scores in attempt $a+1$ to be more than their scores in attempt $a$: 
\vspace{-5pt}
\begin{equation*}
 \mathbold{s}_s\cdot T_{a+1} \cdot \mathbold{q}_{p} - \mathbold{s}_s\cdot T_{a} \cdot \mathbold{q}_{p} \geq 0
\end{equation*}
\vspace{-20pt}

However, this inequality constraint is too strict as the students may occasionally forget the learned concepts (assumption \textbf{(e)}). 
To allow for this occasional forgetting and soften this constraint, we model the knowledge increase as a rank-based constraint, that allows for knowledge loss, but penalizes it.
We formulate this constraint as maximising the value for $\mathcal{L}_2$ in Equation~\ref{eq:obj2}.
Essentially, this penalty term can be viewed as a prediction-consistent regularization.  
It helps to avoid significant changes in students' knowledge level since their performance is expected to transit gradually over time.
\vspace{-10pt}
\begin{equation}
\mathcal{L}_2 = \sum_{j=1}^{a-1} \sum_{s,p} \log \left( \sigma ( \mathbold{s}_s \cdot T_{a} \cdot \mathbold{q}_p -  \mathbold{s}_s \cdot T_{j} \cdot \mathbold{q}_p ) \right)
\label{eq:obj2}
\end{equation}
\vspace{-15pt}
% \begin{equation}
% \mathcal{L}_2 = -\sum_{j=1}^{t-1} \sum_{s,p} \log \left( \sigma ( \mathbold{s}_s \cdot T_{a} \cdot \mathbold{q}_p^{[p]} -  \mathbold{s}_s \cdot T_{j} \cdot \mathbold{q}_p^{[r]} ) \right)
% \label{eq:obj2}
% \end{equation}

%In Equation~\ref{eq:obj2}, the student $s$'s score at $a$ should be higher than the ones of $s$ at $j$ with $j<a$, by rank. 

\noindent\textbf{The Proposed Multi-View Model (MVKM).} We rely on our main assumption \textbf{(i)} to extend our model to capture learning from different learning material types.
So far, we have assumed that course concepts are shared among learning materials (assumption \textbf{(a)}). 
With the knowledge transfer assumption \textbf{(i)}, all learning materials of different types will share the same latent space.
Also, we represent student knowledge and student ability as shared parameters across all different learning material types.
Consequently, for each set of learning materials of type $r \in \mathfrak{R}$, we can rewrite Equation~\ref{eq:simplemodelbias1} as:  
% \vspace{-2pt}
\begin{equation*}
    \hat{x}^{[r]}_{s,a,p} \approx \mathbold{s}_s\cdot T_{a} \cdot \mathbold{q}^{[r]}_p + b_s + b^{[r]}_p + b_a
\end{equation*}
% $$\hat{x}^{[r]}_{s,a,p} \approx \mathbold{s}_s\cdot T_{a} \cdot \mathbold{q}^{[r]}_p + b_s + b^{[r]}_p + b_a$$
An illustration of this decomposition, when considering two learning material types, is presented in Figure~\ref{fig:model}.
Note that we represent one shared matrix student $S$ and one shared knowledge gain tensor $T$ in both types of learning materials.

We can learn the parameters of our model by minimizing the sum of squared differences between the observed (${x}^{[r]}_{s,a,p}$) and estimated ($\hat{x}^{[r]}_{s,a,p}$) values over all learning material types $r \in \mathfrak{R}$.
For the learned parameters to be generalizable to unseen data, we regularize the unconstrained parameters using their L-2 norms.
As a result, we minimize the objective function in Equation~\ref{eq:obj1},  in which $\gamma^{[r]}$ are hyper-parameters that represent the relative importance of different learning materials types. $\lambda_t$ and $\lambda_s$ are hyper-parameters to control the weights of regularization term of $\mathbold{T}$ and $S$.
\vspace{-5pt}

\vspace{-15pt}
\begin{equation}
\begin{split}
\mathscr{L}_1 = \sum_{r,s,a,p}&\gamma^{[r]}(\hat{x}^{[r]}_{s,a,p} -  x^{[r]}_{s,a,p})^2 + \lambda_t \|{T_{a}}\|^2_{F} + \lambda_s \|\mathbold{s}_{s}\|^2_{F}\\
&\text{s.t. } \forall_{r,c,p}\text{ }\ q^{[r]}_{c,p} \geq 0\text{ , } \sum_{c} q^{[r]}_{c,p} = 1
\end{split}
\label{eq:obj1}
\end{equation}

\vspace{-15pt}
Similarly, the knowledge gain and forgetting constraint presented in Equation~\ref{eq:obj2} can be extended to the multi-view model.
Eventually, we use a combination of the reconstruction objective function (Equation~\ref{eq:obj1}) and the learning and forgetting objective function (Equation~\ref{eq:obj2}) to model students' knowledge increase, while representing their personalized knowledge and finding learning material latent features, as in Equation~\ref{eq:obj3}.
Note that, since our goal is to minimize $\mathscr{L}_1$ and maximize $\mathscr{L}_2$, we use $-\mathscr{L}_2$ to minimize $\mathscr{L}$.
To balance between the accuracy of student performance prediction and modeling student knowledge increase, we use a nonnegative trade-off parameter $\omega$:
\vspace{-5pt}
\begin{equation}
\mathscr{L} = \mathscr{L}_1 - \omega \mathscr{L}_2
\label{eq:obj3}
\vspace{-10pt}
\end{equation}

% \noindent\textbf{Learning Model Parameters}
We use stochastic gradient descent algorithm to minimize $\mathscr{L}$ in Equation~\ref{eq:obj3}. The parameters need to learn are students' latent feature matrix ($S$), dynamic knowledge in each concept at any attempt ($\mathbold{T}$), importance of each concept in every learning material (${Q}^{[r]}$), each student's general ability ($b_s$), each learning material's difficulty/helpfulness ($b^{[r]}_p$), and each attempt's bias ($b^{[r]}_a$).

\section{Experiments}
\label{sec:experiments}
%To evaluate our model, we run three kinds of experiments.
We evaluate our model with three sets of experiments.
First, to validate if the model captures the variability of observed data, we use it to predict unobserved student performances (Sec.~\ref{sec:experiments:performance}).
Second, to check if our model represents valid student knowledge growth, we study the knowledge increase patterns between different types of students and across different concepts (Sec.~\ref{sec:experiments:knowledge}). 
% \textcolor{red}{we should add learning resource concept analysis}
Finally, to study if the model meaningfully recovers learning materials' latent concepts, we analyze their similarities according to the learned latent feature vectors (Sec.~\ref{sec:experiments:resource}). 
Without loss of generalizability, although the model is designed to handle multiple learning material types, we experiment on two learning material types.
Before the experiments, we will go over our datasets, and experiment setup. 
\begin{table}[!ht]
\centering
% \scalebox{0.8}{
\resizebox{0.48\textwidth}{!}{
\begin{tabular}{|c|c|c|c|c|c|c|}
\hline
Dataset & \begin{tabular}[c]{@{}l@{}}material\\ type 1 (\#)\end{tabular} & \begin{tabular}[c]{@{}l@{}}material\\ type 2 (\#)\end{tabular} & \#stu & \begin{tabular}[c]{@{}l@{}}act.\\ seq. \\len.\end{tabular} & \#rcds. & \begin{tabular}[c]{@{}l@{}}avg. \\  sco.\end{tabular}\\ 
\hline\textbf{}
Synthetic\_NG & quiz (10) & discussion (15) & 1000 & 20 & 19991 & 0.6230 \\
Synthetic\_NG2 & quiz (10) & discussion (15) & 1000 & 20 & 19991 & 0.6984 \\
Synthetic\_G & quiz (10) & assignment (15) & 1000 & 20 & 19980 & 0.6255   \\
\hline
MORF\_QD & assignment (18) & discussion (525) & 459  & 25 & 6800 & 0.8693 \\
MORF\_QL & assignment (10) & lecture (52) & 1329 & 76 & 58956 & 0.7731 \\
% \hline
% Canvas\_BM & quiz (31)                                                      & discussion (35)                                                & 480       & 30                                                        & 0.9068                                                   \\
Canvas\_H & quiz (10) & discussion (43) & 1091 & 20 & 13633 & 0.8648 \\ \hline               
\end{tabular}}
\caption{Statistics for each datasets, where \#stu is number of students, act. seq. len. is the maximum activity length, \#rcds. is number of records that student interact with learning materials and avg. sco. is graded learning material's average score.}
\label{Table:genstats}
\vspace{-12pt}
\end{table}
\subsection{Datasets}
\begin{figure*}[ht]
\vspace{-5pt}
\centering
% \subfigure[Synthetic datasets]{
\subfigure{
\includegraphics[width=0.25\textwidth]{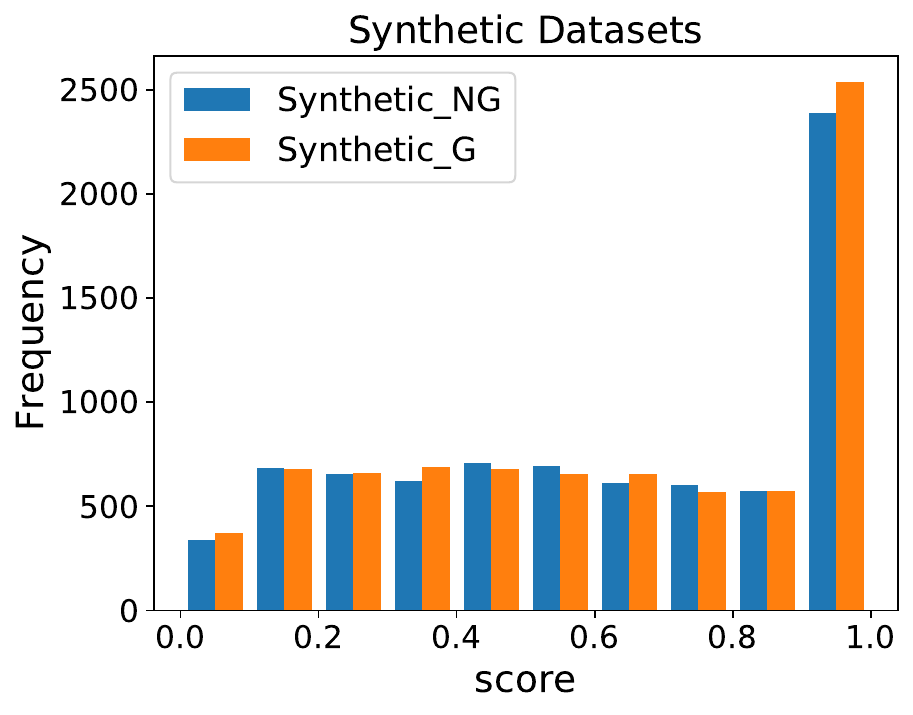}
}
\hspace{-10pt}
% \subfigure[Synthetic datasets]{
\subfigure{
\includegraphics[width=0.25\textwidth]{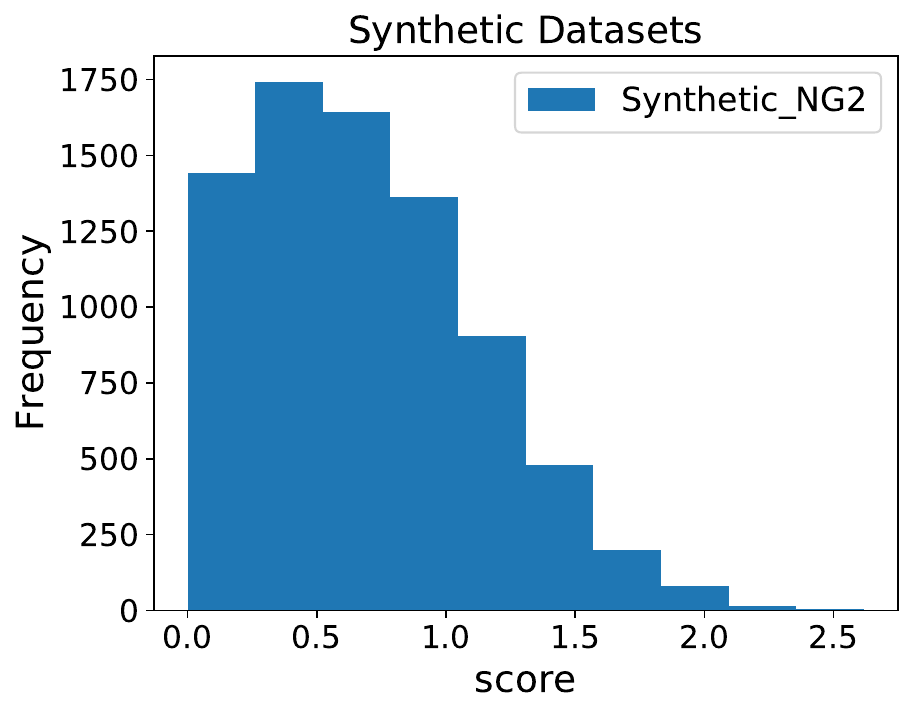}
}
% \vspace{-25pt}
\hspace{-10pt}
% \subfigure[Real-World datasets]{
\subfigure{
\includegraphics[width=0.25\textwidth]{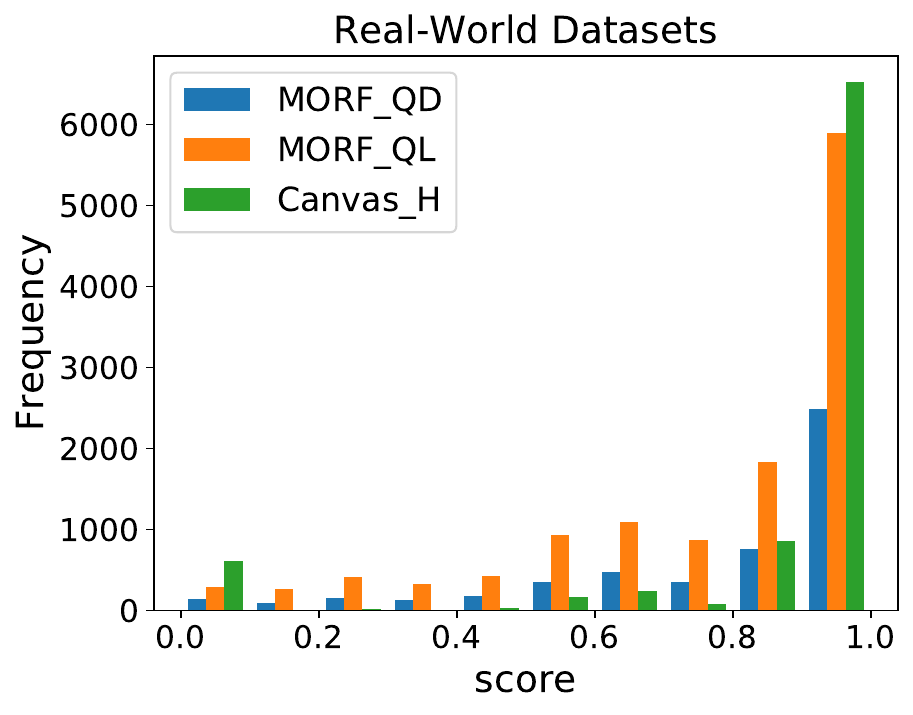}
}
\vspace{-15pt}
\caption{Histogram of graded materials' Scores in Synthetic Data and Real-World Data.}
\label{fig:hists}
\vspace{-10pt}
\end{figure*}

We use three synthetic and three real-world datasets (from two MOOCs) to evaluate the proposed model.
Our choice of real-world datasets is guided by two factors, aligned with our assumptions: that they include multiple types of learning material, and that they allow the students to work freely with the learning material in the order they choose.
In the real-world datasets, we select the students that work with both types of learning materials, removing the learning materials that none of these students have interacted with.
General statistics of each dataset are presented in Table~\ref{Table:genstats}. Figure~\ref{fig:hists} shows score distributions of the graded learning material types in these datasets.

% \begin{figure*}[ht]
% \centering
% \includegraphics[width=0.9\textwidth]{pics_croped/hist.pdf}

% \caption{Histogram of graded materials' Scores in Synthetic Data and Real-World Data.}
% \label{fig:hists}
% \end{figure*}
\noindent\textbf{Synthetic Data. }We generate three synthetic datasets according to two characteristics: (1) if both learning material types are graded vs. if one of them is non-graded (or has binary observations); (2) if the student scores are capped and their distribution is highly skewed vs. if the score distribution in not capped and less skewed.
%kinds of synthetic datasets: (1) when the two learning material types are graded with a score, and (2) when the first type is graded but the second type is non-graded (or has binary observations). 
% \textcolor{red}{so that the dataset we generate take both graded and non-graded learning material into account.}

% \textcolor{red}{For creating dataset that both material are graded, take two types of learning materials as the example, we generate synthetic data follow the similar assumptions made by our model.}
For creating the datasets, we follow similar assumptions as to the ones made by our model.
Expecting $P^{[1]}$ learning materials of type $1$, and $P^{[2]}$ materials of type $2$, %and assuming $a_s$ as the interaction sequence of student $s$ with the learning material, 
we first generate a random sequence $L_s$ for each student $s$, which represents the student's attempts on different learning materials. 
Considering $C$ latent concepts, we then create two random matrices $Q^{[1]} \in \mathbb{R}^{C \times P^{[1]}}$ and $Q^{[2]} \in \mathbb{R}^{C \times P^{[2]}}$ as the mapping between the learning material and the $C$ underlying concepts, such that the sum of values for each underlying learning material is one. 
% we then normalize these two matrices to make sure that for each learning material, 
% Next, we generate a $N \times K$ random matrix as the mapping between $N$ students and the $K$ latent student features, denote as $S$.
% As initial matrix for relationship between students latent feature and concepts, we generate a random $K \times C$ matrix $\mathcal{T}_{1}$. %TODO: how to define
For the student knowledge gain assumption, we represent each student's knowledge increase separately.
Hence, we directly create a student knowledge tensor $\mathbold{K}$, instead of creating $S$ and $\mathbold{T}$, and multiplying them.
To generate $\mathbold{K}$, we first generate a random matrix $K_{1}$ that represents all students' initial knowledge in all $C$ concepts.
For generating the knowledge matrix in the next attempts ($K_{a}$), we use the following random process. 
For each student $s$, we generate a random number $\alpha$ representing the probability of forgetting. If $\alpha > \theta$ (forgetting threshold), we assume no forgetting happens and increase the value in the knowledge matrix according to the learning material that the student has interacted with: $\mathbold{k}_{s, a} = \mathbold{k}_{s,a-1} + \beta \mathbold{q}^{[r]}_{L_s[a]}$. 
Here, $\beta$ is a random effect of increasing and $L_s[a]$ is the learning material that student has selected to interact with at timestamp $a$.
Otherwise ($\alpha < \theta$, or forget), we set $k_{s, a, c} = k_{s, a-1, c} - \text{rand}(0,\epsilon)$, for $\forall c \in C$.
we use n-mode tensor product to build $\mathbold{X}^{[1]}$ and $\mathbold{X}^{[2]}$, where $\mathbold{X}^{[1]} = \mathbold{K} Q^{[1]}$, $\mathbold{X}^{[2]} = \mathbold{K} Q^{[2]}$. 
Finally, according to the student learning sequences $L_s$, we remove the ``unobserved'' values that are not in $L_s$ from $\mathbold{X}^{[1]}$ and $\mathbold{X}^{[2]}$.

To create different data types according to the first characteristic above, for the graded learning material type $r$, we keep the values in $\mathbold{X}^{[r]}$. 
For the non-graded ones, we use the same process as above, except that in the final step we set $x^{[r]}_{s, a, p}=1$ according to the student sequence $L_s$. 
% This process creates a nearly balanced distribution of grades for students that has a relatively wide range (Figure~\ref{fig:hists}).
However, in many real-world scenarios, the score distribution of students is highly skewed especially towards higher scores (Figure~\ref{fig:hists} show it). 
To represent this skewness, in some of the generated datasets, we clip all $x^{[r]}_{s,a,p} > 1$ to 1.

Then, we create following three datasets according to above process: \textit{Synthetic\_G}, in which both learning material types are graded and scores are skewed; \textit{Synthetic\_NG}, in which one of the learning material types is graded and scores are skewed; and \textit{Synthetic\_NG2}, in which one of the learning material types is graded and scores are not skewed. 
% We generate all synthetic data with 1000 students, $P^{[1]} = 10$ learning materials of type 1, $P^{[2]} = 15$ learning materials of type 2, $C = 3$ latent concepts, and maximum sequence length of 20 for students.
We generate all synthetic data with 1000 students, $P^{[1]} = 10$ learning materials of type 1, $P^{[2]} = 15$ learning materials of type 2, $C = 3$ latent concepts, and maximum sequence length of 20 for students.

%In the following experiment setting, we employ two 
% thresholded
% datasets that we clip all $x^{[r]}_{s,a,p} > 1$ to 1, one has two graded learning materials and the other one has one non-graded learning material one graded learning material. We denote this two dataset as Synthetic\_G and Synthetic\_NG.
% We also evaluate on one original dataset that we do not make values of $x^{[r]}_{s,a,p} > 1$ to 1, for this dataset, we create one graded learning material and one non-graded learning material, we denote this dataset as Synthetic\_NG2.

%As the reason that in real-world dataset, student normalized grade are always not larger than one. Then, we clip all $x^{[r]}_{s,a,p} > 1$ to 1.

% In the following experiment setting, we denote the Synthetic data has two graded learning material as Synthetic\_QA, the other synthetic dataset with only one graded material as Synthetic\_QD.}
% \textcolor{red}{add student feature and concept dimension when create simulation data, change synthetic data to TQ}

% \begin{figure*}[h]
%     \centering
%     \includegraphics[width= 1\textwidth]{pics/prediction_step.pdf}
%     % \includegraphics[scale=0.46]{pics/eval.pdf}
%     \caption{Evaluation on a target student sequence: given the student's first half of attempt sequence, we will evaluate the predicted scores on all quizzes in the rest of sequence iteratively. Taking First (left) and second(right) iteration of online testing case as an example to illustrate the evaluation process.}
%     \label{fig:prediction step}
% \end{figure*}

\noindent\textbf{Canvas Network~\cite{Canvas-Network2016}. }
This is an online available dataset collected from various courses on the Canvas network platform~\footnote{\url{http://canvas.net}}.
The available open online course data comes from various study fields, such as computer science, business and management, and humanities. 
For each course, its general field of study is presented in the data.
The rest of the dataset is anonymized such that course names, discussion contents, student IDs, submission contents, or course contents are not available.
Each course can have different learning material types, including assignments, discussions, and quizzes.
We experiment on the data from one course in this system, with course id $770000832960975$, which is in the humanities field (Canvas\_H dataset). 
We use quizzes as the graded learning material type and discussions as the non-graded one.
% We experiment on the data from two courses in this system, namely course id $770000832945397$ and $770000832960975$. 
% The first course is in the business and management (Canvas\_BM) and the second course is in the humanities (Canvas\_H) field.
% We use quizzes as the graded learning material type and discussions as the non-graded ones for these datasets.

\noindent\textbf{MORF~\cite{Andres2016}. }
This is a dataset of the ``educational data mining'' course~\cite{ryanbigdata} at Coursera\footnote{\url{https://www.coursera.org/}}, available via the MOOC Replication Framework (MORF). %We use a dataset for a course on the topic of educational data mining. 
The course includes various learning material types, including video lectures, assignments, and discussion forums.
Students' history, in terms of their watched video lectures, submitted assignments, and participated discussions, in addition to the score they received in assignments, is available in data.
In this course, we experiment with two datasets, each focusing on two sets of learning material types: one with assignments as the graded type and discussions as the non-graded type (MORF\_QD), another with assignments as the graded type and video lecture views as the non-graded type (MORF\_QL).
% \vspace{0pt}
\vspace{-5pt}
\subsection{Experiment Setup}
We use $5$-fold student-stratified cross-validation to separate our datasets into test and train.
At each fold, we use interaction records from $80\%$ of students as training data. 
For the rest ($20\%$) of the students (target students), we split their attempt sequences on the graded learning material type into two parts: the first $50\%$ and the last $50\%$.
% \textcolor{red}{For each quiz, we predict each target student's performance in their first attempt on that quiz, that happens in the last $50\%$ of their sequence. 
% Predicting first attempts is a common practice in the educational data mining literature~\cite{Baker2014}.}
For performance prediction experiments, we predict the performance of the graded learning material type in the last $50\%$, given the first $50\%$. 
In order to see how the proposed model captures the knowledge growth, we do online testing, in which we predict the test data attempt by attempt (next attempt prediction). 
%Particularly, we first predict the test students' scores in the first attempt of the test data, then put use this test data to update the model again for predicting the next attempt. We repeat these steps until the last attempt. 
% An illustration of data splitting and online testing are shown in figure \ref{fig:prediction step}. 
% -Eventually, we report the average performance on all five folds.For selecting the best hyper-parameters, such as number of concepts or regularization coefficients, we use a separate validation dataset.
Eventually, we report the average performance on all five folds.
For selecting the best hyper-parameters, we use a separate validation dataset. Our code and synthetic data are available at GitHub\footnote{\href{https://github.com/sz612866/MVKM-Multiview-Tensor}{https://github.com/sz612866/MVKM-Multiview-Tensor}}.

% \vspace{-30pt}

% \subsubsection{5-fold Cross Validation}
% \begin{enumerate}
%     \item Shuffle the whole dataset and randomly split it into 5 groups
%     \item For each groups
%     \begin{enumerate}
%         \item Take the group as a hold out or test data set
%         \item Take the remaining groups as a training data set and validation set
%         \item Fit a model on the training set, tune the hyper-parameters on validation set, and evaluate it on the test set
%         \item Retain the evaluation score and discard the model
%     \end{enumerate}
%     \item report the averaged performance on 5 test data set
% \end{enumerate}

\subsection{Student Performance Prediction}
\vspace{-2pt}
\label{sec:experiments:performance}
In this set of experiments, we test our model on predicting student scores on their future unobserved graded learning material attempts.
More specifically, we estimate student scores on their future attempts, and compare them with their actual scores in the test data.
\vspace{-5pt}
\subsubsection{Baselines}
We compare our model with state-of-the-art student performance prediction baselines:\\
\textbf{Individualized Bayesian Knowledge Tracing (IBKT)}
~\cite{johnsonscaling, Yudelson2013}: This is a variant of the standard BKT model, which assumes binary observations and provides individualization on student priors, learning rate, guess, and slip parameters~\footnote{The code is from \url{https://github.com/CAHLR/pyBKT}}.\\
\textbf{Deep Knowledge Tracing (DKT) \cite{Piech2015}}: DKT is a pioneer algorithm that uses recurrent neural networks to model student learning, %which reports substantial improvements in prediction performance 
on binary (success/failure) student scores.\\
\textbf{Feedback-Driven Tensor Factorization (FDTF)}~\cite{Sahebi2016}: This tensor factorization model decomposes the student interaction tensor into a learning material latent matrix and a knowledge tensor. However, it only models one type of learning material, does not capture student latent features, and does not allow the students to forget the learned concepts. It assumes that students' knowledge strictly increases as they interact with learning materials.\\
\textbf{Tensor Factorization Without Learning} (TFWL): This is a model similar as FDTF, the only difference is TFWL does not have constraint that student knowledge is increasing.\\ 
\textbf{Rank-Based Tensor Factorization (RBTF)}~\cite{Doan2019a}: This model has similar assumptions to FDTF. Except, it allows for occasional forgetting of concepts and has extra bias terms. Compared to MVKM, it does not differentiate between different student groups. It only uses student previous scores in graded learning materials to predict students' future scores, and it has a different tensor factorization strategy.\\
\textbf{Bayesian Probabilistic Tensor Factorization (BPTF)}
~\cite{Xiong2010}: This is a recommender systems model has a smoothing assumption over student scores in consecutive attempts.\\
\textbf{AVG}: This baseline uses the average of all students' scores for all predictions.

As mentioned before, one major issue in real-world datasets is their skewness, meaning that, on average, student grades are skewed towards a full (complete) score on quizzes/assign-ments. 
This skewness adds to the complexity of predicting an accurate score for unobserved quizzes: only using an overall average score will provide a relatively good estimate of the real score. 
As a result, outperforming a simple average baseline is a challenging task. 
% -and it is essential to add AVG as a baseline.

The mentioned baselines all work on one type of learning material. 
Since our proposed MVKM model works with more than one learning material type, to be fair in evaluations, we run baseline algorithms in a multi-view setup.
To do this, we aggregate the data from all learning material types and use that as an input to these baselines. %to build a large student s feed these algorithms the aggregated data on all learning material types.
In those cases, we add a ``MV'' to the end of their names.
For example, FDTF\_MV represents running FDTF on aggregation of student interactions with multiple learning material types.
In addition, for knowledge tracing algorithms (BKT and DKT) which are designed for binary student responses (correct or incorrect), we modify their settings to make them predict numerical scores as described below. 
First, we binarize students' historical scores based on median score. 
Specifically, if the score is greater than the median, it will be set to 1, and 0 otherwise.
Then, we use the probability of success generated by BKT and DKT as the probability of student receiving a score more than median score. 
Eventually, the numerical predicted scores can be obtained by viewing the probability output as the percentile of students' score on that specific question. 
Moreover, since these models require pre-defined knowledge components (KCs), we assume that each learning material is mapped to one KC in these models.
%Then the outputs of both BKT and DKT models are probabilities of being greater than median score, and the numerical predicted scores can be obtained by viewing the probability output as the percentile of student's historical score on that specific question. 

In addition to the above, we compare our multi-view model with its basic variation (MVKM-Base) using the data from graded materials only, and its multi-view variation without the learning and forgetting constraints (MVKM-W/O-P).

\begin{table*}[!ht]
\vspace{-5pt}
\centering
\resizebox{0.9\textwidth}{!}{
\begin{tabular}{|c|c|c|c|c|c|c|}
\hline
\multirow{2}{*}{Methods} & \multicolumn{2}{c|}{Synthetic\_NG}                                                        & \multicolumn{2}{c|}{Synthetic\_NG2}                                                       & \multicolumn{2}{c|}{Synthetic\_G}                                                        \\ \cline{2-7} 
                         & RMSE                                        & MAE                                         & RMSE                                        & MAE                                         & RMSE                                        & MAE                                         \\ \hline
AVG                      & 0.3084$\pm$0.0072          & 0.2820$\pm$0.0093          & 0.5059$\pm$0.0115          & 0.4005$\pm$0.0115          & 0.3070$\pm$0.0039          & 0.2811$\pm$0.0050          \\
RBTF                     & 0.2515$\pm$0.0126          & 0.2027$\pm$0.0081          & 0.3374$\pm$0.0234          & 0.2681$\pm$0.0146          & 0.2628$\pm$0.0113          & 0.2103$\pm$0.0080          \\
FDTF                     & 0.4906$\pm$0.0172            & 0.4410$\pm$0.0207        & 0.6588$\pm$0.0215          & 0.5529$\pm$0.0226          &            0.5041$\pm$0.0184        & 0.4537$\pm$0.0213              \\
TFWL                  & 0.5283$\pm$0.0168            & 0.4632$\pm$0.0178          & 0.6919$\pm$0.0132          & 0.5883$\pm$0.0156         & 0.5490$\pm$0.0053          & 0.5130$\pm$0.0076          \\
BPTF                     & 0.1675$\pm$0.0048          & 0.1256$\pm$0.0061          & 0.3454$\pm$0.0140          & 0.2589$\pm$0.0072          & 0.1825$\pm$0.0064          & 0.1381$\pm$0.0050          \\
IBKT                      & 0.4744$\pm$0.0118          & 0.4197$\pm$0.0140          & 0.6630$\pm$0.0122          & 0.5494$\pm$0.0152          & 0.4748$\pm$0.0076          & 0.4233$\pm$0.0098          \\
DKT                      & 0.2694$\pm$0.0275          & 0.1911$\pm$0.0241          & 0.4536$\pm$0.0404          & 0.3569$\pm$0.0413          &  0.2716$\pm$0.0209        & 0.2047$\pm$0.0178           \\\hline
RBTF-MV                  & 0.2920$\pm$0.0069          & 0.2305$\pm$0.0078          & 0.4064$\pm$0.0213          & 0.3227$\pm$0.0147          & 0.2618$\pm$0.0155          & 0.2126$\pm$0.0130          \\
FDTF-MV                  & 0.4078$\pm$0.0168          & 0.3402$\pm$0.0167           & 0.5861$\pm$0.0211          & 0.4688$\pm$0.0135         & 0.4888$\pm$0.0112          & 0.4538$\pm$0.0131          \\
TFWL-MV                  & 0.4337$\pm$0.0139         & 0.3896$\pm$0.0133          & 0.6386$\pm$0.0161          & 0.5450$\pm$0.0194         & 0.5312$\pm$0.0137          & 0.4626$\pm$0.0145          \\
BPTF-MV                  & 0.1718$\pm$0.0037          & 0.1457$\pm$0.0055          & 0.3438$\pm$0.0158          & 0.2603$\pm$0.0120          & 0.1533$\pm$0.0055          & 0.1184$\pm$0.0044          \\
IBKT-MV                   & 0.4257$\pm$0.0142          & 0.3585$\pm$0.0155          & 0.6019$\pm$0.0124          & 0.4892$\pm$0.0165          & 0.4844$\pm$0.0068          & 0.4275$\pm$0.0089          \\
DKT-MV                   & 0.4278$\pm$0.0313          & 0.3613$\pm$0.0318          & 0.6399$\pm$0.0515          & 0.5320$\pm$0.0526          & 0.3390$\pm$0.0252          & 0.2892$\pm$0.0245          \\\hline
MVKM-Base               & 0.2007$\pm$0.1069          & 0.1498$\pm$0.0809          & 0.3026$\pm$0.0697          & 0.2273$\pm$0.0356          & 0.2097$\pm$0.0485          & 0.1565$\pm$0.0348          \\
MVKM-W/O-P               & 0.1714$\pm$0.0089          & 0.1306$\pm$0.0089          & 0.2817$\pm$0.0316          & 0.2213$\pm$0.0245          & 0.1796$\pm$0.0345          & 0.1357$\pm$0.0190          \\
Our Method (MVKM)        & \textbf{0.1388$\pm$0.0048} & \textbf{0.1049$\pm$0.0056} & \textbf{0.2221$\pm$0.0074} & \textbf{0.1739$\pm$0.0048} & \textbf{0.1532$\pm$0.0128} & \textbf{0.1171$\pm$0.0097}\\\hline
\end{tabular}
}
\caption{Performance Prediction results on synthetic datasets, measured by RMSE and MAE, shown with variance in 5-fold cross-validation}
\label{table:synthetic_results}
\end{table*}

\begin{table*}[!]
\centering
\resizebox{0.9\textwidth}{!}{
\begin{tabular}{|c|c|c|c|c|c|c|}
\hline
\multirow{2}{*}{Methods} & \multicolumn{2}{c|}{MORF\_QD}                                                       & \multicolumn{2}{c|}{MORF\_QL}                                                     & \multicolumn{2}{c|}{CANVAS\_H}                                                    \\ \cline{2-7} 
                         & RMSE                                      & MAE                                     & RMSE                                    & MAE                                     & RMSE                                    & MAE                                     \\ \hline
AVG                      & 0.2410$\pm$0.0227            & 0.1913$\pm$0.0161          & 0.2420$\pm$0.0108          & 0.1957$\pm$0.0067          & 0.0767$\pm$0.0121          & 0.0555$\pm$0.0040          \\
RBTF                     & 0.2711$\pm$0.0229            & 0.2132$\pm$0.0147          & 0.2572$\pm$0.0114          & 0.1980$\pm$0.0074          & 0.1571$\pm$0.0172          & 0.1235$\pm$0.0103          \\
FDTF                     & 0.3081$\pm$0.0437            & 0.2401$\pm$0.0329          & 0.3006$\pm$0.0194          & 0.2324$\pm$0.0151          & 
0.1395$\pm$0.0259        & 0.0929$\pm$0.0119                \\
TFWL                     & 0.2750$\pm$0.0529           & 0.2003$\pm$0.0249          & 0.3090$\pm$0.3090          & 0.2237$\pm$0.0099            &
0.2377$\pm$0.0803        & 0.1186$\pm$0.0513          \\
BPTF                     & 0.2172$\pm$0.0128            & 0.1776$\pm$0.0082          & 0.2302$\pm$0.0068          & 0.1953$\pm$0.0048          & 0.1114$\pm$0.0120          & 0.0946$\pm$0.0082          \\
IBKT                      & 0.2756$\pm$0.0070            & 0.2281$\pm$0.0053          & 0.2646$\pm$0.0147          & 0.2174$\pm$0.0096          & 0.0856$\pm$0.0105          & 0.0692$\pm$0.0042          \\
DKT                      & 0.3169$\pm$0.0374            & 0.2498$\pm$0.0313          & 0.2859$\pm$0.0061          & 0.2158$\pm$0.0075          & 0.0911$\pm$0.0322          & 0.0616$\pm$0.0173          \\ \hline
RBTF-MV                  & 0.2814$\pm$0.0282            & 0.2177$\pm$0.0222          & 0.2624$\pm$0.0193          & 0.1977$\pm$0.0136          & 0.1484$\pm$0.0098          & 0.1171$\pm$0.0054          \\
FDTF-MV                  & 0.3138$\pm$0.0441            & 0.2453$\pm$0.0387         & 0.2398$\pm$0.0137           & 0.1866$\pm$0.0091          & 
0.1149$\pm$0.0085        & 0.0907$\pm$0.0068               \\
TFWL-MV                  & 0.2919$\pm$0.0275           & 0.1975$\pm$0.0160          & 0.3222$\pm$0.0208          & 0.2178$\pm$0.0165           & 
0.1748$\pm$0.0600        & 0.0784$\pm$0.0269          \\
BPTF-MV                  & 0.2615$\pm$0.0129            & 0.2286$\pm$0.0114          & 0.2313$\pm$0.0070          & 0.1960$\pm$0.0041          & 0.1452$\pm$0.0100          & 0.1343$\pm$0.0081          \\
IBKT-MV                   & 0.2774$\pm$0.0204            & 0.2177$\pm$0.0099          & 0.2904$\pm$0.0098          & 0.2137$\pm$0.0062          & 0.0834$\pm$0.0125          & 0.0425$\pm$0.0049          \\
DKT-MV                   & 0.2938$\pm$0.0310            & 0.2352$\pm$0.0236          & 0.2540$\pm$0.0065          & 0.2185$\pm$0.0047          & 0.079$\pm$0.0247           & 0.0496$\pm$0.0065          \\ \hline
MVKM-Base               & 0.2242$\pm$0.0328            & 0.1669$\pm$0.0207          & 0.2277$\pm$0.0119          & 0.1724$\pm$0.0081          & 0.0666$\pm$0.0159          & 0.0411$\pm$0.0040          \\
MVKM-W/O-P               & 0.2385$\pm$0.0196                  & 0.1771$\pm$0.0104              & 0.2450$\pm$ 0.0145               & 0.1814$\pm$0.009          & 0.0649$\pm$0.0111          & 0.0388$\pm$0.0027                \\
Our Method (MVKM)        & \textbf{0.2088 $\pm$ 0.0229} & \textbf{0.1603$\pm$0.0142} & \textbf{0.2150$\pm$0.0127} & \textbf{0.1654$\pm$0.0104} & \textbf{0.0613$\pm$0.0112} & \textbf{0.0362$\pm$0.0028} \\ \hline
\end{tabular}
}
\caption{Performance Prediction results on real-world datasets, measured by RMSE and MAE, shown with variance in 5-fold cross-validation.}\label{table:real_results}
\label{table: real data results}
\vspace{-10pt}
\end{table*}

\subsubsection{Performance Metrics and Comparison}
In this task, our target is to accurately estimate the actual student scores.
To evaluate how close our predicted values are to the actual ones, we use Root Mean Squared Error (RMSE) and Mean Absolute Error (MAE) between the predicted scores and the actual scores for students.
% \textcolor{blue}{In order to compare the performance with BKT and DKT, both of which were designed for the binarized response prediction problem, we design the experiment on binarized student response prediction with performance metric area under curve (AUC).}
Table \ref{table:synthetic_results} and \ref{table:real_results} show the results of performance among different methods on synthetic data and real data, respectively. 
We can see that our proposed model outperforms other baselines on synthetic data, and has the best performance on real datasets in general.
% Moreover, we also validate that the performance of our model without additional learning resource (this means we just use one learning material to capture student feature and dynamic knowledge tensor), and we denote this as method MVKM-Base.

\noindent\textbf{MVKM-Base vs. Single Material Type Baselines. }Comparing MVKM-Base with other algorithms that use student scores only, shows us that MVKM-Base has consistently lower error compared to most baselines, in both synthetic and real-world datasets.
This result demonstrates the ability of MVKM-Base in capturing data variance and validity of its assumptions for real-world graded data. 
Compared to AVG, MVKM-Base can represent more variability; compared to RBTF, the student latent features in MVKM-Base leads to improved results; compared to FTDF, the forgetting factor results in less error; and compared to BKT and DKT, modeling the learning material concepts in $Q$ and having a rank-based constraint to enforce learning improves the performance.
The only baseline algorithm that outperforms MVKM-Base in some setups is BPTF. 
Particularly, BPTF has a lower RMSE and MAE in Synthetic\_NG and Synthetic\_G datasets that are skewed. 
In real-world datasets, it performs better than MVKM-Base in MORF-QD dataset that is more sparse and has a slightly higher average score compared to the other two.
This shows that BPTF is better than MVKM-Base in handling skewed data. 
One potential reason is BPTF's smoothing assumption, in contrast with MVKM-Base's rank-based knowledge increase, that results in a more homogeneous score predictions for each student.  

\noindent\textbf{MVKM: Multiple vs. Single Material Types. }Comparing MVKM's results with MVKM-Base model, we can see that using data from multiple learning material types improves performance prediction results.
It verifies our assumptions regarding knowledge transfer in different learning material types through the knowledge gain in shared concept latent space.
This is given that in other models, e.g., all models except DKT in MORF-QD, adding different learning material types increases the prediction error.
This error increase is particularly happening with BPTF model in real-world datasets and DKT model in synthetic ones.
This shows that merely aggregating data from various resources, without appropriate modeling, can even harm the prediction results.
This difference between MVKM and other baselines is in its specific setup, in which each learning material type is modeled separately, while keeping a shared knowledge space, student latent features, and knowledge gain.

\noindent\textbf{Learning and Forgetting Effect. }To further test the effect of our knowledge gain and forgetting constraint, we compare MVKM with MVKM-W/O-P, a variation of our proposed model without the rank-based constraint in Equation~\ref{eq:obj2}.
We can see that MVKM outperforms MVKM-W/O-P in all datasets. 
This shows that the soft knowledge increase and forgetting assumption is essential in correctly capturing the variability in students' learning.
Particularly, comparing MVKM-W/O-P's results with MVKM-Base, the single-view version that includes the rank-based learning constraints, we can measure the effect of adding multiple learning material types vs. the effect of adding the learning and forgetting constraints in MVKM model.
In CANVAS\_H dataset, adding multiple learning material types is more effective than learning constraint, and in MORF datasets, realizing learning constraint is more important than modeling multiple types of learning materials.
Nevertheless, they are not mutually exclusive and both are important in the model.

%Moreover, we also validate that the performance of MVKM-Base model, this means we just use one learning material to capture student feature and dynamic knowledge tensor. 
%In order to see the rank-based condition of our model is really meaningful, we validate our model without this condition as well (it means we ignore penalty term in our model), and we denote it as method MVKM-W/O-P. The results show that both MVKM-Base and MVKM-W/O-P perform worse than our proposed model.

% \begin{figure*}[]
% \centering
% \hspace{-80pt}
% \subfigure[MORF\_QD]{
% \includegraphics[width=.40\textwidth]{pics/MORF-QD-hyperparameters.pdf}
% }
% \hspace{-30pt}
% \subfigure[MORF\_QL]{
% \includegraphics[width=.40\textwidth]{pics/MORF-QL-hyperparameters.pdf}
% }
% \hspace{-30pt}
% \subfigure[CANVAS\_H]{
% \includegraphics[width=.40\textwidth]{pics/CANVAS-H-hyperparameters.pdf}
% }
% \hspace{-70pt}
% \caption{Sensitivity Analysis on Hyper-Parameters: Penalty Weight and Markovian Step on MORF-QD, MORF-QL, and CANVAS-H Datasets}
% \label{fig:hyperparameters}
% \end{figure*}

\noindent\textbf{Hyper-parameter Tuning}
% There is some way to set a good value for the parameter.
% The exact value of the parameter makes little difference.
Using a separate validation set, we experiment with various values (grid search) for model hyper-parameters to select the most representative ones for our data. 
% The hyper-parameters that we need to tune are: student latent feature dimension $K$, question latent feature dimension $C$, penalty weight $\omega$, importance of different learning resources  $\gamma^{[r]}$, learning rate $\eta$, Markovian steps $m$ and regularization parameter $\lambda_t$. 
% regularization parameters $\lambda_1, \lambda_2, \lambda^{[r]}_3, \lambda_b$. 
% Also, we need to tune the regularization parameters $\lambda_t$ and $\lambda_b$, for which we choose one identical regularization parameter $\lambda$ in the experiments. 
Specifically, we first vary the student latent feature dimension $K$ in $[1, 5, 10 ,\cdots, 40, 45]$, the question latent feature dimension $C$ in $[1,2,\cdots,9,10]$, the penalty weight $\omega$ in $[0.01, 0.05, 0.1, 0.5,$
$1, 2, 3]$, the Markovian step $m$ in $[1,2,\cdots, 10]$, and the learning resource importance parameter $\gamma^{[r]}$ in $[0.05, 0.1, 0.2, 0.5,$
$1, 2]$. 
Once we found a good set of hyper-parameters from coarse-grained grid search, we search the values close to the optimal values to find out the best fine-grained values for these hyper-parameters. 
The best resulting hyper-parameter values for each dataset are listed in table \ref{table:hyperparameters}. 
We use $\gamma^{[1]}$ as the trade-off parameter for graded learning material, $\gamma^{[2]}$ for anther learning material.
As we can see, in both the synthetic and real-world data, the learning and forgetting constraint is more important (larger $\omega$) when having a non-graded learning material type. 
This shows that binary interaction data, unlike student grades (or scores), is not precise enough to represent the students' gradual knowledge gain in the absence of a learning and forgetting constraint.
Also, comparing $\gamma^{[2]}$ in MORF\_QD vs. MORF\_QL we can see that the importance of video lectures is more than discussions in predicting students' performance.

\begin{table}[!h]
\centering
\resizebox{0.475\textwidth}{!}{
\begin{tabular}{|c|c|c|c|c|c|c|c|c|c|}
\hline
Dataset        & K  & C & $\omega$   & $\gamma^{[1]}$ & $\gamma^{[2]}$   & $\eta$    & m & $\lambda_t$ &$\lambda_s$ \\\hline
Synthetic\_NG  & 3  & 3 & 0.2 & 1  & 0.1  & 0.1  & 1 & 0.01 & 0.001\\
Synthetic\_NG2 & 3  & 3 & 0.2 & 1  & 0.1  & 0.1  & 1 & 0.001 & 0.001\\
Synthetic\_G   & 3  & 3 & 0.1 & 1  & 0.4  & 0.1  & 1 & 0.001 & 0.001\\\hline
MORF\_QD       & 39 & 5 & 1   & 1  & 0.05 & 0.1  & 1 & 0     & 0\\
MORF\_QL       & 35 & 9 & 0.6 & 1  & 0.5  & 0.1  & 1 & 0     & 0\\
Canvas\_H      & 28 & 7 & 2.0 & 1  & 0.5  & 0.01 & 1 & 0    & 0\\\hline
\end{tabular}
}
\caption{Hyperparameters of our model for each dataset}
\label{table:hyperparameters}
\vspace{-0pt}
\end{table}

% We also demonstrate the sensitivity of two hyper-parameters, penalty weight and Markovian steps, in figure \ref{fig:hyperparameters}. For better visualization, we only show the performance with penalty weight in $[0.1, 0.2, \cdots, 1.0]$ and Markovian steps in $[1, 2, \cdots, 5]$. The performance for either penalty weight or Markovian step equals to $0$ is the performance of method MVKM-W/O-P listed in the table \ref{table:real_results}. 

% \begin{table}[H]
% \centering
% \resizebox{0.48\textwidth}{!}{
% \begin{tabular}{|c|c|c|c|c|c|c|c|}
% \hline
% Dataset                        & \begin{tabular}[c]{@{}l@{}}K\end{tabular} & \begin{tabular}[c]{@{}l@{}}C\end{tabular} & $\omega$ & \begin{tabular}[c]{@{}l@{}}$\gamma$ \end{tabular} & \begin{tabular}[c]{@{}l@{}}$\eta$ \end{tabular} & $m$ & $\lambda$  \\ 
% \hline\textbf{}
% Synthetic\_NG & 3 & 3 & 0.2 &0.1 & 0.1 & 1 & 0.01 \\
% Synthetic\_NG2 &3  &3 &0.2 &0.1 & 0.1 & 1 & 0.001 \\
% Synthetic\_G &3  &3 &0.1 &0.4  &0.1 & 1 & 0.001 \\
% \hline
% MORF\_QD &39  &5  &1.0 &0.05 &0.1 & 1  & 0 \\
% MORF\_QL &35  &9 &0.6 &0.5 &0.1 & 1 & 0  \\
% Canvas\_H &28 &7 &2.0 &0.5 &0.01 & 1& 0 \\ \hline                                                  
% % \hline
% % Canvas\_BM & quiz (31)                                                      & discussion (35)                                                & 480       & 30                                                        & 0.9068                                                   \\

% \end{tabular}}
% \caption{Hyperparameters of Our Model for Each Dataset}
% \label{table:hyperparameters}
% \end{table}

\vspace{-2pt}
\subsection{Student Knowledge Modeling}
\label{sec:experiments:knowledge}
In this set of experiments, we answer two main research questions: 1) Can our model's learning and forgetting constraint capture meaningful knowledge trends across concepts for students as a whole? and 2) Are the individual student's knowledge growth representative of their learning? To answer these questions, we look at the estimated knowledge tensor of students ($\mathbold{K} = S\mathbold{T}$).
%There are 2 main questions we would like to ask for the ability of student knowledge modeling in our proposed model. First, could our model capture meaningful knowledge dynamic for students in overall ? Second, could our model capture interpretable knowledge growth individually, which differentiate among students ?
 % grows as they interact with learning materials and if this growth is varied for different concepts.
% \begin{figure}[!h]
% \centering
% % \vspace{-15pt}
% % \hspace{-50pt}
% % \subfigure[MORF\_QD]{
% % \includegraphics[width=.37\textwidth]{pics/MORF-QD-concept_growth_comparison.pdf}
% % }
% % \hspace{-25pt}
% \subfigure[MORF\_QL]{
% \includegraphics[width=.3\textwidth]{pics_croped/MORF-QL-concept_growth_comparison.pdf}
% }
% % \hspace{-25pt}
% \subfigure[CANVAS\_H]{
% \includegraphics[width=.3\textwidth]{pics_croped/CANVAS-H-concept_growth_comparison.pdf}
% }
% % \hspace{-60pt}
% \caption{Average knowledge gain of each concept across all students in three different courses. }
% \label{fig:conAvgStd}
% \end{figure}

\begin{figure}[!h]
\centering
\vspace{-5pt}
% \hspace{-50pt}
% \subfigure[MORF\_QD]{
% \includegraphics[width=.37\textwidth]{pics/MORF-QD-concept_growth_comparison.pdf}
% }
% \hspace{-25pt}
\includegraphics[width=.47\textwidth]{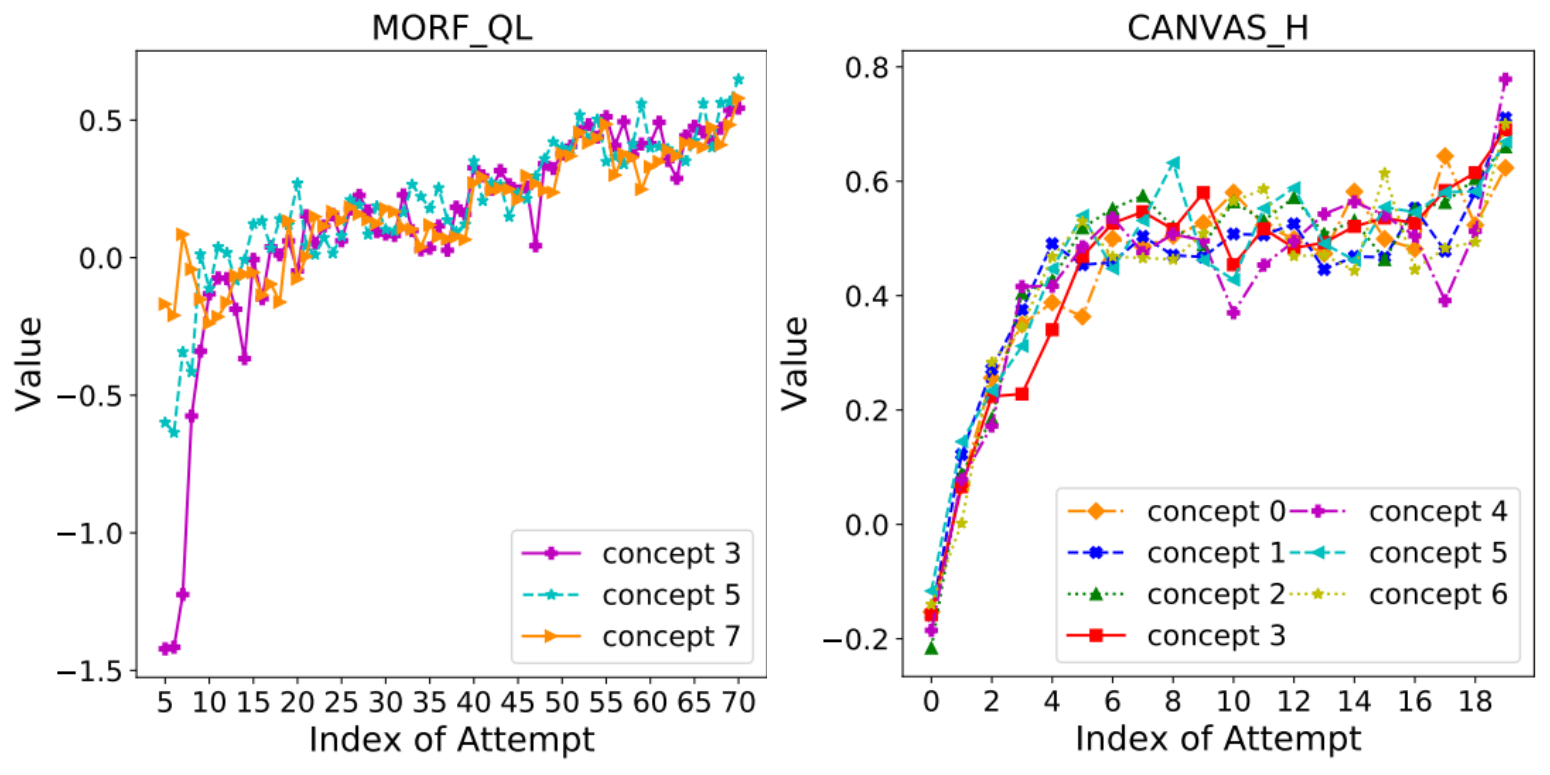}
\vspace{-10pt}
% \hspace{-60pt}
\caption{Average knowledge gain of concepts across all students.}
\label{fig:conAvgStd}
\vspace{-5pt}
\end{figure}
To answer the first question, we check the average student knowledge growth on different concepts. Figure~\ref{fig:conAvgStd} shows the average knowledge of all students in different concepts (represented with different colors) during the whole course period (X-axis) for MORF\_QL, and CANVAS\_H datasets (MORF\_QD has similar patterns as MORF\_QL, we don't show it due to the page limitation). Notice that, for a clear visualization, we only show $3$ out of $9$ concepts from MORF\_QL dataset in the figure.
We can see that, on average, students' knowledge in different concepts increase.
% This increase is varied among the concepts in some datasets, such as MORF\_QL, while being more homogeneous in others, such as in MORF\_QD.
Particularly, in MORF\_QL, the initial average knowledge on concept 3 is less that concepts 5 and 7.
However, students learn this concept rapidly as shown by the increase of knowledge level around the tenth attempt.
As the knowledge growth is less smooth in this concept, compared to the other two (e.g., the drop around the $15^{th}$ attempt), students are more likely to forget it rapidly. 
Eventually, the average student knowledge in all concepts are close to each other.
On the other hand, in CANVAS\_H, the average initial knowledge in different concepts are relatively close. 
However, students end up having different knowledge levels in different concepts at the end of the course, especially in concepts 0 and 4.
Also, all six concepts show large fluctuations across the attempts. 
%As you can see, our proposed model is capable to monitor knowledge acquisition level of each concept at each time stamp. 
Overall, the students have a significant knowledge gain at the first few attempts and the knowledge gain slows down after that. 
This is aligned with our expectation on students' knowledge acquisition through out the course. 
% \begin{figure}[!ht]
% \centering
% % \vspace{-15pt}
% % \hspace{-50pt}
% % \subfigure[MORF\_QD]{
% % \includegraphics[width=.37\textwidth]{pics/MORF-QD-concept_growth_comparison_without_penalty.pdf}
% % }
% % \hspace{-25pt}
% \subfigure[MORF\_QL]{
% \includegraphics[width=.35\textwidth]{pics/MORF-QL-concept_growth_comparison_without_penalty.pdf}
% }
% % \hspace{-25pt}
% \subfigure[CANVAS\_H]{
% \includegraphics[width=.35\textwidth]{pics/CANVAS-H-concept_growth_comparison_without_penalty.pdf}
% }
% % \hspace{-60pt}
% \caption{Without penalty term, average knowledge gain of each concept across all students in three different courses is counter-intuitive.}
% \label{fig:conAvgStdNoPenalty}
% \end{figure}

\begin{figure}[!ht]
\centering
% \vspace{-15pt}
% \hspace{-50pt}
% \subfigure[MORF\_QD]{
% \includegraphics[width=.37\textwidth]{pics/MORF-QD-concept_growth_comparison_without_penalty.pdf}
% }
% \hspace{-25pt}
\includegraphics[width=.47\textwidth]{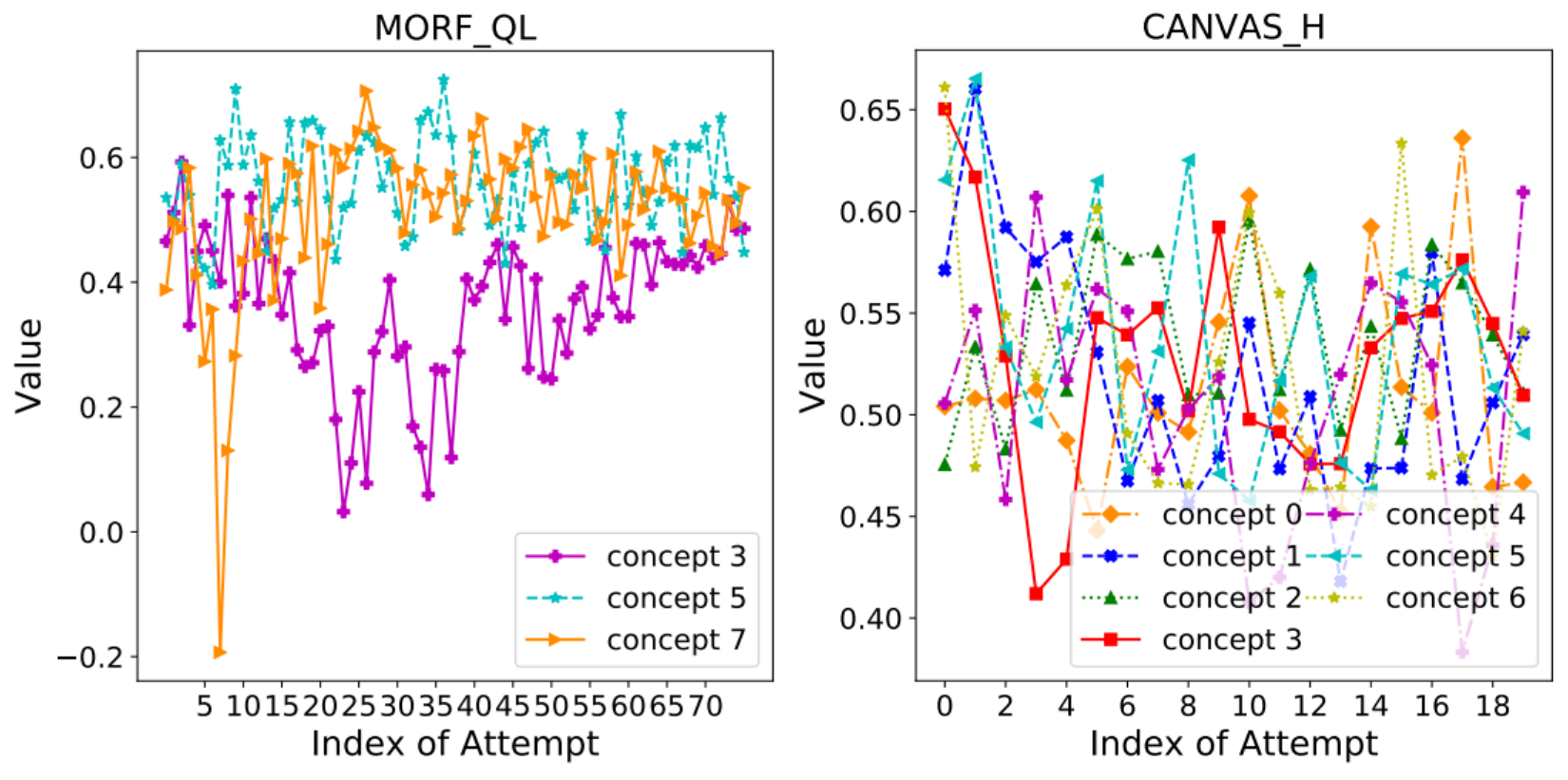}
\vspace{-10pt}
% \hspace{-60pt}
% \caption{Without penalty term, average knowledge gain of each concept across all students in three different courses is counter-intuitive.}
\caption{Average knowledge gain of each concept across all students.}
\label{fig:conAvgStdNoPenalty}
% \vspace{-10pt}
\end{figure}

To show the effect of the learning and forgetting constraint in MVKM, we look at the student knowledge acquisition in the MVKM-W/O-P model, that removes this constraint.
The MVKM-W/O-P's average student knowledge in different concepts throughout all attempts is shown in Figure~\ref{fig:conAvgStdNoPenalty}. 
We can see that despite its acceptable performance prediction error, MVKM-W/O-P's estimated knowledge trends are elusive and counter-intuitive.
% For example, many concepts (such as concept 0 in MORF\_QD and concept 3 in MORF\_QL) show a U-shaped curve.
For example, many concepts (such as concept 3 in MORF\_QL) show a U-shaped curve.
This curve can be interpreted as the students having a high prior knowledge in these concepts, but forgetting them in the middle of the course, and then re-learning them at the end of the course.
In some cases, such as concept 1 in CANVAS\_H, students lose some knowledge and forget what they already knew, by the end of the course.
This demonstrates the necessity of learning and forgetting penalty term in MVKM.

For second question, we check if there are meaningful differences between knowledge gain trends of different students. 
To do this, we apply spectral clustering on students' latent features matrix $S$ to discover different groups of students. 
Then, we compare students' learning curves from different clusters. 
The number of clusters is determined by the significance of difference on average performance in each cluster. 
We obtained 3 clusters of students for $MORF\_QD$ course, and 2 clusters for $MORF\_QL$ and $CANVAS\_H$ courses based on students' latent features from our model.

\begin{figure}[!h]
\centering
% \vspace{-5pt}
\includegraphics[width=.47\textwidth]{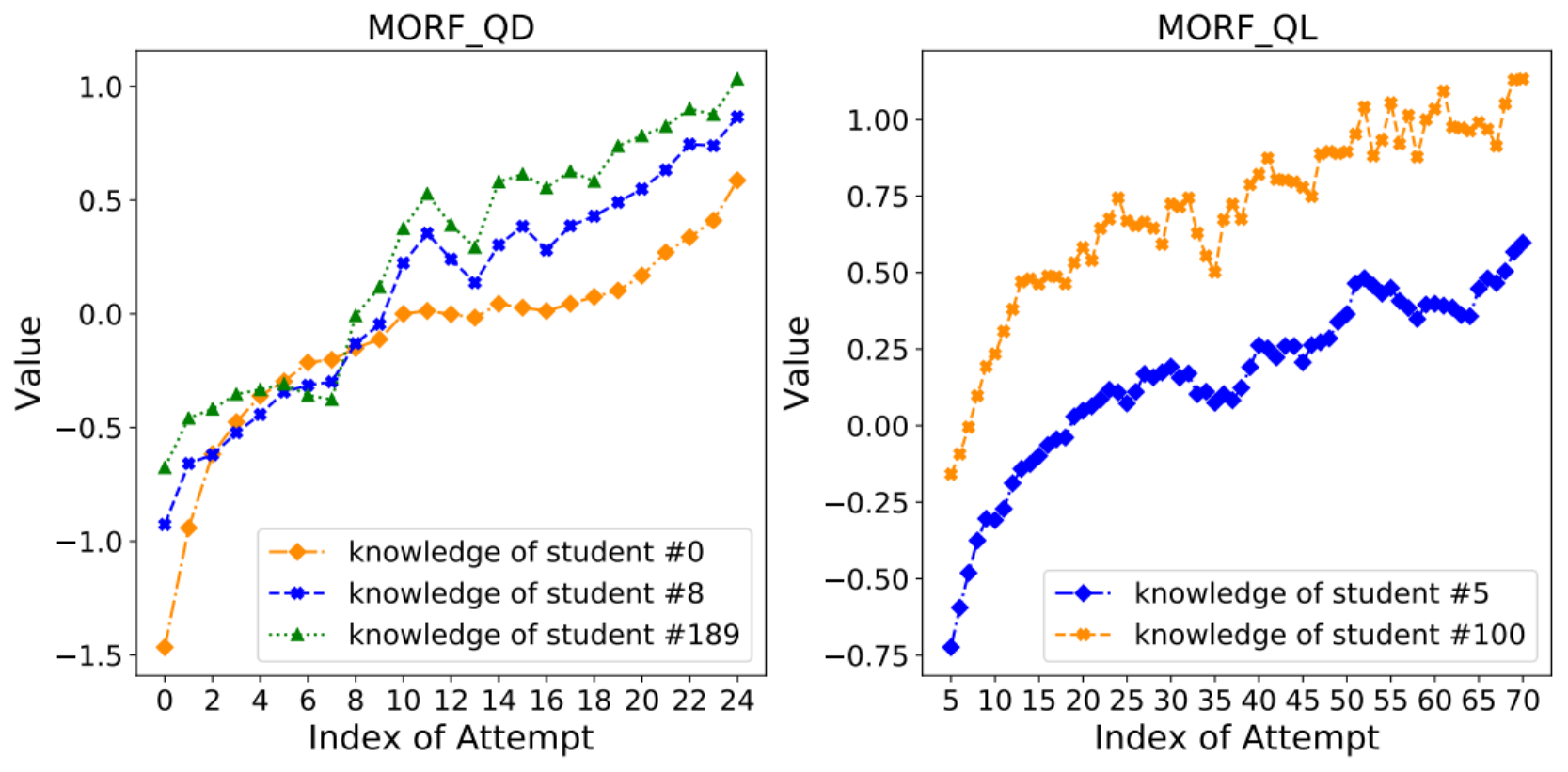}
\vspace{-10pt}
\caption{Sample students' knowledge gain across all concepts in two different courses.}
% In MORF\_QD, student $\#0$, $\#8$, and $\#189$ have average grades $0.202, 0.636$, and $0.909$ respectively, which align with the knowledge levels shown in the figure. Student $\#5$ and $\#100$ in MORF\_QL have average grades $0.9$ and $0.98$ respectively.In addition, we found that student $\#100$ finished much more questions than student $\#5$. Student $\#0$ and $\#3$ in CANVAS\_H have average grades $0.883$ and $0.69$ respectively. }
\label{fig:3stdAvgCon}
\vspace{-12pt}
\end{figure}

To see the differences in these groups, we sample one student from each cluster in each real-world dataset. 
% Figure~\ref{fig:3stdAvgCon} shows these sample students' knowledge gain, averaged over all concepts, in datasets MORF\_QD, MORF\_QL, and CANVAS\_H. 
Figure~\ref{fig:3stdAvgCon} shows these sample students' knowledge gain, averaged over all concepts, in datasets MORF\_QD and MORF\_QL (CANVAS\_H is not showed due to the page limitaion, it has similar patterns as MORF\_QD). 
The figures show that different students start with different initial prior knowledge.
For example, in MORF\_QL, student $\#5$ starts with a lower prior knowledge than student $\#100$ and ends up with a lower final knowledge.
Also, the figure shows that different knowledge gain trends across students, particularly in MORF\_QD. 
For example, student $\#0$ starts with a lower prior knowledge than the other two students, but has a faster knowledge growth, and catches up with them around attempt $8$. 
However, this student's knowledge growth slows down after a while end up to be lower than the other two at the end of course.
To see if the quantified knowledge is meaningful, we compare student's knowledge growth with their scores. Students $\#0$, $\#8$, and $\#189$ in MORF\_QD have average grades $0.202, 0.636$, and $0.909$, in MORF\_QL, $\#5$ and $\#100$ have average grades $0.9$ and $0.98$. This align with the knowledge levels shown in the figure.
% -For example, looking at scores of students $\#0$, $\#8$, and $\#189$ for MORF\_QD, they have average grades $0.202, 0.636$, and $0.909$ respectively, which align with the knowledge levels shown in the figure. In MORF\_QL, $\#5$ and $\#100$ have average grades $0.9$ and $0.98$, and moreover student $\#100$ finished much more questions than student $\#5$.
% We find the same pattern in other datasets (explained in Figure~\ref{fig:3stdAvgCon}'s caption). 
%TODO: do we have a weak student crossing a strong one?
% \textcolor{red}{In both figures, student A starts with a higher initial knowledge and continues to grow, and student C starts with a low initial knowledge and also grows. 
% Student B in CANVAS\_H course starts with the same initial knowledge as student C, but differentiates from it after the second attempt.}
%  In CANVAS\_H, student $\#0$'s average score ($0.883$) is higher than student $3$'s (). 
%  The reason student $0$ and $5$ belong to two different clusters is student 0 spent less attempts on each question, which indicate higher knowledge acquisition ability than student $5$.
% Looking at scores of students A, B, and C, we can see that in both datasets, student A's average scores ($0.95$ in CANVAS\_H and $0.785$ in MORF\_QD) are higher than student B's ($0.799$ in CANVAS\_H and $0.648$ in MORF\_QD), and student B's average scores are higher than student C's ($0.6875$ in CANVAS\_H and $0.202$ in MORF\_QD).
% The difference between learned knowledge and average scores of students B and C in MORF\_QD is more than that of students A and B. 
These observations show that MVKM can meaningfully differentiate between different students' knowledge growth. 

% \textcolor{red}{To further evaluate this, we calculate the auto-correlation between knowledge growth and student scores} %TODO

% \textcolor{blue}{To further validate the necessity of penalty term (eq. \ref{eq:obj2}) and additional learning resource, we first compare the performance with and without the penalty term, and then compare the knowledge growth pattern with and without the penalty term on real data. We can see that in tables \ref{table:synthetic_results} and \ref{table:real_results}, penalty term indeed improve the performance on both synthetic data and real data. In addition, \ref{} }

% \begin{figure}
%     \centering
%     \begin{subfigure}[t]{0.20\textwidth}
%         \centering
%         \includegraphics[width=\linewidth]{pics/CANVAS-H-knowledges_growth_comparison_over_students.pdf}
%         \caption{}
%       \label{fig:}
%     \end{subfigure}
%     \hspace{12pt}
%     \begin{subfigure}[t]{0.20\textwidth}
%         \centering
%         \includegraphics[width=\linewidth]{pics/MORF-QD-knowledge_growth_comparison_over_students.pdf}
%         \caption{}
%         \label{fig:}
%     \end{subfigure}
%     \caption{}
%     \label{fig:}
% \end{figure}

\subsection{Learning Resource Modeling}
\label{sec:experiments:resource}
% We experimented with single/group/complete linkage, but found this choice made little difference, we therefore report only
% to study if the model meaningfully recovers learn-ing materials’ latent concepts, we analyze their similarities accord-ing to the learned latent feature vectors (Section 4.5)
In this section, we evaluate our model on how well it can represent the variability and similarity of different learning materials.
We mainly focus on two questions:
1) Are the learning materials' biases consistent with their difficulty levels? 
2) Are the discovered latent concepts for learning materials (matrix $Q^{[r]}$) representative of actual conceptual groupings of learning materials in the real datasets?

% \begin{figure*}[!ht]
% \centering
% \subfigure[MORF\_QD]{
% \includegraphics[width=.35\textwidth]{pics/MORF-QD-avg_score_and_bias_q_correlation.pdf}
% }
% \hspace{-28pt}
% \subfigure[MORF\_QL]{
% \includegraphics[width=.35\textwidth]{pics/MORF-QL-avg_score_and_bias_q_correlation.pdf}
% }
% \hspace{-28pt}
% \subfigure[CANVAS\_H]{
% \includegraphics[width=.35\textwidth]{pics/CANVAS-H-avg_score_and_bias_q_correlation.pdf}
% }
% \caption{Graded learning materials' average score v.s. bias of learning materials learned by MVKM. The spearman correlation on MORF\_QD is 0.779, on MORF\_QL  is 0.597, on CANVAS\_H is 0.960.}
% % \caption{Graded learning materials' average score v.s. bias of learning materials learned by MVKM. The spearman correlation on MORF\_QD is 0.779 with p-value 0.0001, on MORF\_QL  is 0.597 with p-value 0.073, on CANVAS\_H is 0.960 with p-value 0.00001.}
% \label{fig:bias_q_correlation}
% \end{figure*}

\noindent\textbf{Bias Evaluation. }For the first question, since we do not have access to the learning materials' difficulty levels, we use average student scores on them, as a proxy for difficulty.
As a result, we only use graded learning materials for this analysis. 
% We compare the question bias captured by our model with the average score of each question. 
% We plot a figures to visualize (See Figure \ref{fig:bias_q_correlation}). 
We calculated the spearman correlation between question bias captured by our model and average score of each question. The spearman correlation on MORF\_QD is 0.779, on MORF\_QL  is 0.597, and on CANVAS\_H is 0.960.We find that question bias derived from MCKM is highly correlated with average question score, where the lower the actual average grades are, the lower the bias values are learned. 

% -We calculated the spearman correlation between question bias captured by our model and average score of each question. The spearman correlation on MORF\_QD is 0.779 with p-value 0.0001, on MORF\_QL  is 0.597 with p-value 0.073, and on CANVAS\_H is 0.960 with p-value 0.00001.
% -We find that question bias derived from our model is highly correlated with average question score, where the lower the actual average grades are, the lower the bias values are learned. This also validates the effectiveness of our proposed model. 

\begin{figure*}[!ht]
\vspace{-7pt}
\centering
\subfigure[]{
\includegraphics[width = .475\textwidth]{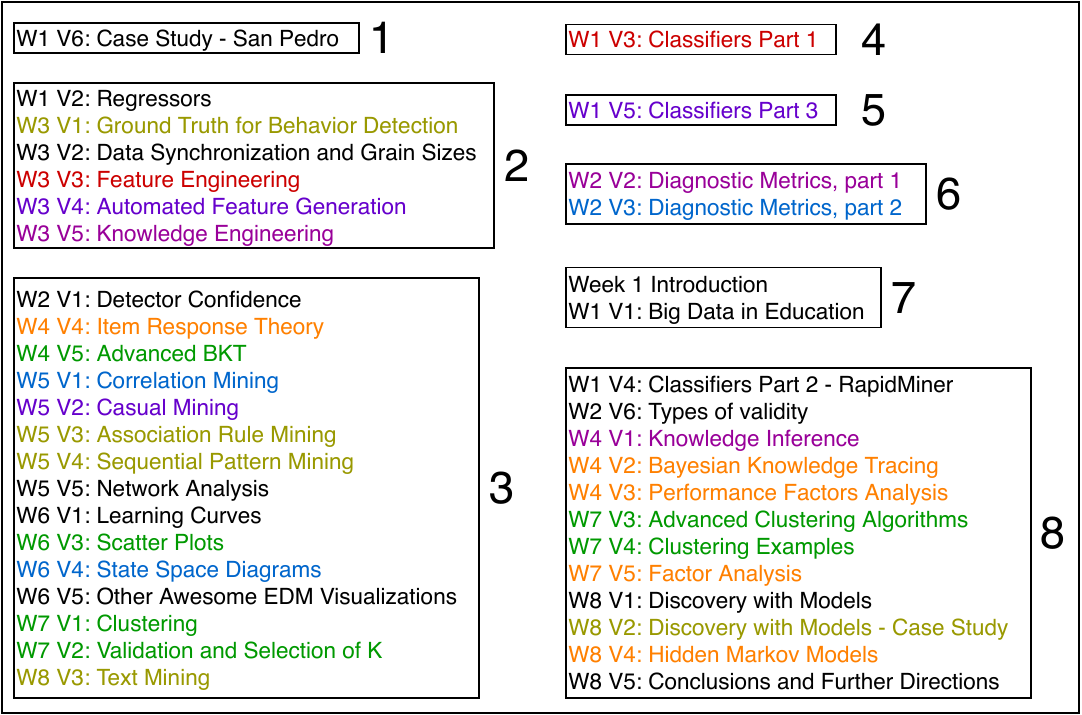}
\label{fig:clusters without quiz our}}
\vspace{-5pt}
\subfigure[]{
\includegraphics[width = .475\textwidth]{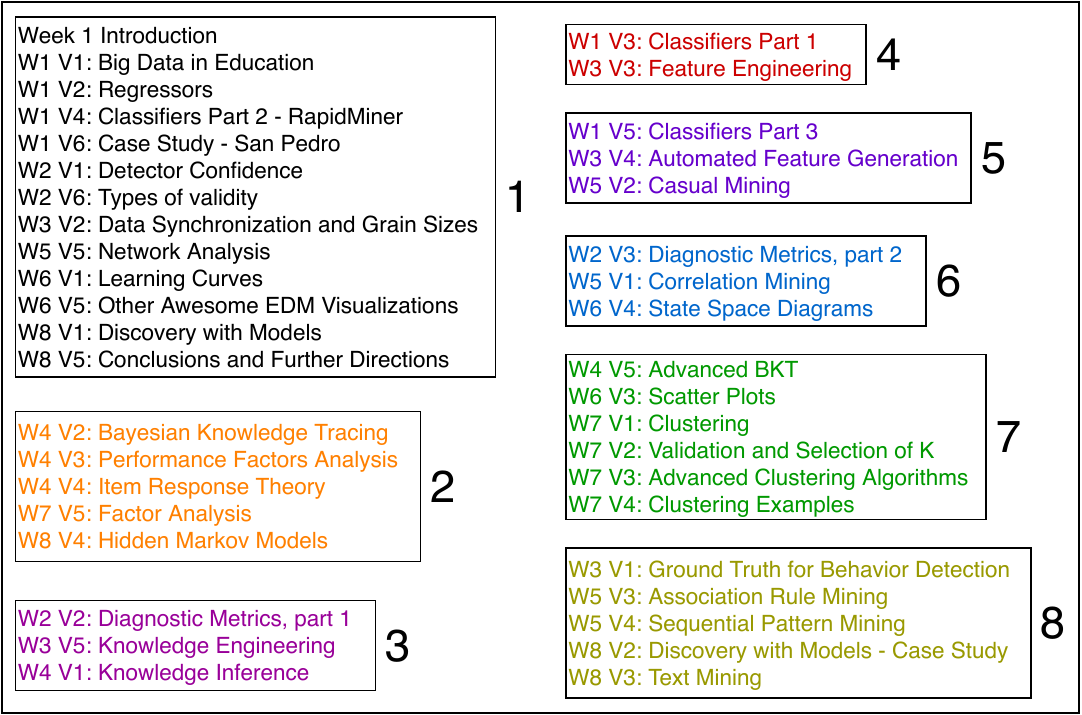}
\label{fig:clusters without quiz LDA}}
\vspace{-10pt}
\caption{Clusters that were discovered by using MVKM (a), clusters discovered by using video-lecture transcripts (b).}
\label{fig:clusters without quizzes}
\vspace{-10pt}
\end{figure*}

\begin{figure*}[!ht]
\centering
\subfigure[]{
\includegraphics[width = .475\textwidth]{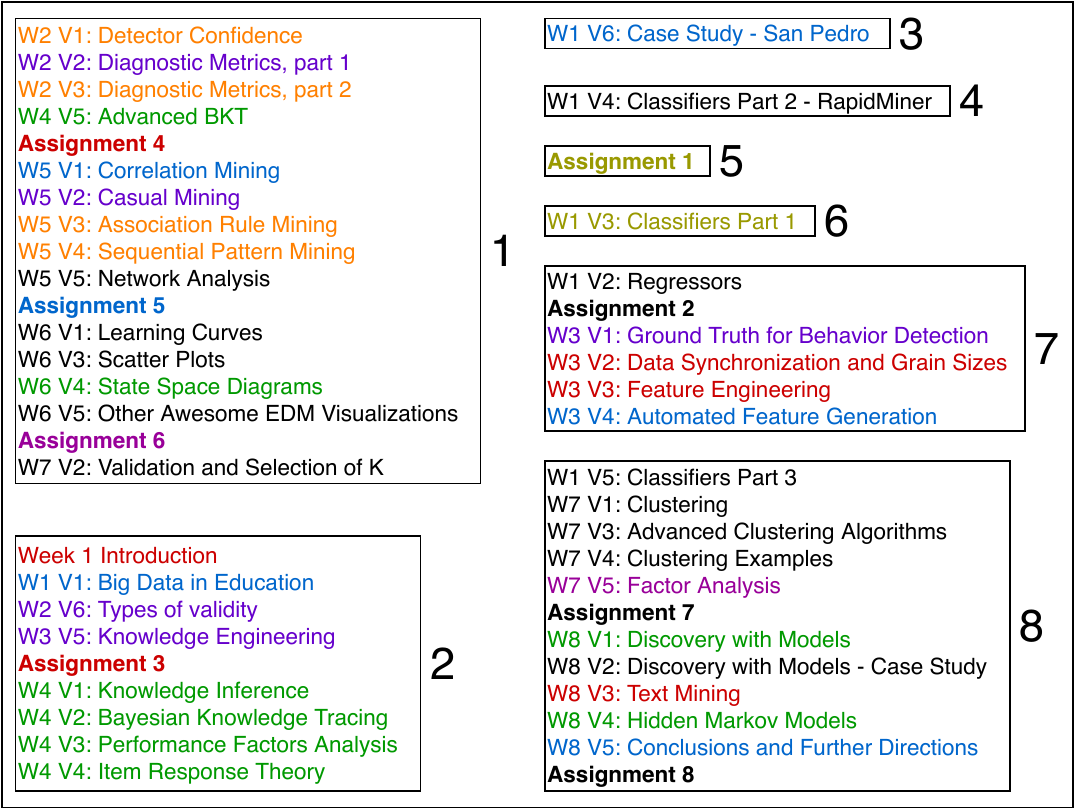}
\label{fig:clusters with quiz our}
}
\vspace{-5pt}
\subfigure[]{
\includegraphics[width = .475\textwidth]{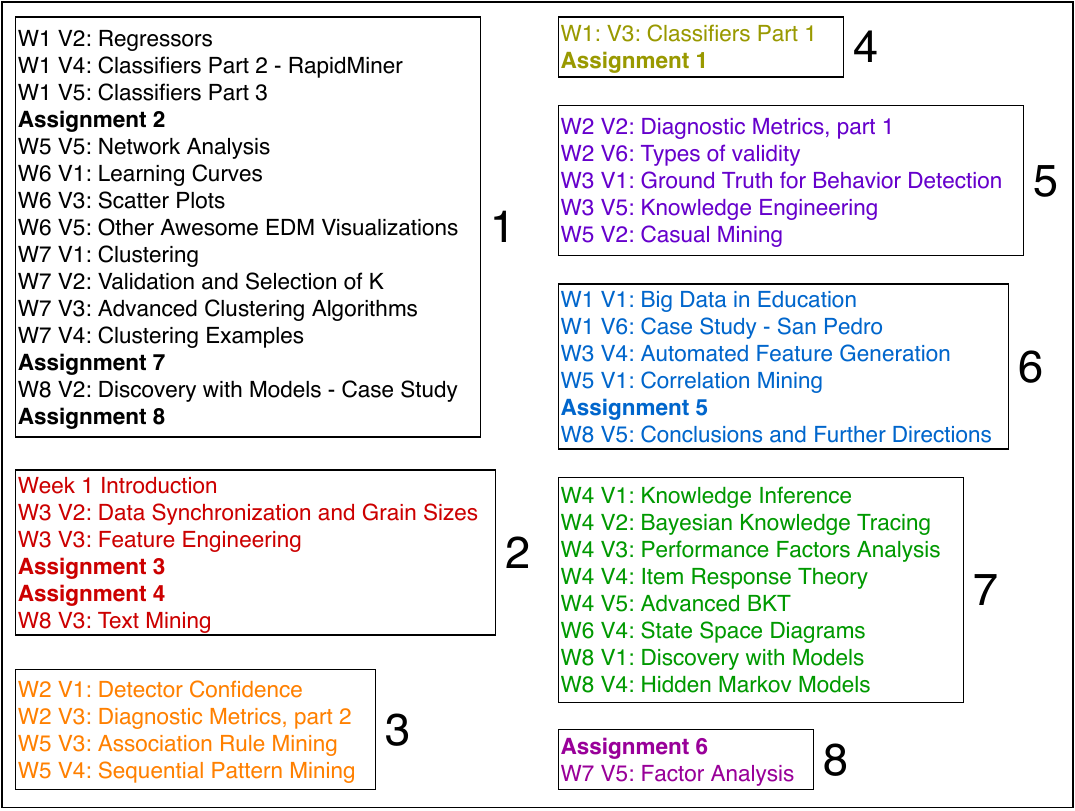}
\label{fig:clusters with quiz LDA}
}
\vspace{-10pt}
\caption{Clusters discovered by using MVKM (a), clusters discovered by using video-lecture transcripts and assignment texts(b).}
% \caption{MORF lecture videos' and quizzes' clustering result by using $Q^{[r]}$ of MVKM (left) and content information (right).}
\label{fig:clusters with quizzes}
\vspace{-15pt}
\end{figure*}

\noindent\textbf{Within-Type Concept Evaluation. }For the second question, we would like to know how much the learning materials' discovered latent concepts resemble the real-world similarities in them.
To evaluate the real-world similarities, we rely on two scenarios: 1) the learning material that are arranged closely to each other in the course structure, either in same module or in consequent modules, are similar to each other (course structure similarity); 2) the learning materials that are similar to each other have similar concepts and contents (content similarity). 
Since only one of our real-world datasets, MORF\_QL, includes the required information for these scenarios, we use this dataset in the continuation of this paper. 
For first scenario, the course includes an ordered list modules, each of which include an ordered list of videos, in addition to the assignments associated with each module. 

For the second scenario, because our learning materials are not labeled with their concepts in our datasets, we use their textual contents (not used in MVKM) as a representation of their concepts.
Particularly, we have subscripts for 40 video lectures, and text of questions for 8 quizzes.
We note that if two learning materials present the same concepts, their textual contents should also be similar to each other. 
As a result, we build content-based clusters of learning materials, each of which containing the learning materials that are conceptually similar to each other. 
Specifically, to cluster the learning material according to their contents, we use Spectral Clustering on the latent topics that are discovered using Latent Dirichlet Analysis (LDA)\cite{blei2003latent} on the learning material's textual contents.
%Due to the lack of information of Canvas dataset, we don't know any learning materials' relative features such as content, topics or title. It is not valid to evaluate our learning rescource modeling. 
%We rely on MORF dataset where we have partial quizzes' content and transcripts and titles of lecture videos to evaluate the second question.
%We only evaluate on MORF\_QL dataset, the reason is that lecture videos' transcripts is less noise compare with students' discussions who has some topics or terms are not relative to concepts. There are 40 lectures and 8 quizzes that we can find their lecture video transcript or quizzes content. 
In the same way, we can cluster the learning materials according to their discovered latent concepts by MVKM.
Similar to the textual analysis, we use spectral clustering on the discovered $Q^{[r]}$ matrices to form clusters of learning materials.
% Then, we compare and contrast the discovered clusters using MVKM with the two scenarios stated above.
To do this, we first consider only one learning material type (the video lectures) and then move on to the similarities between two types of learning materials (both video lectures and assignments).

The results are shown in Figure~\ref{fig:clusters without quizzes} for within-type learning material similarity in video-lectures.
%We firstly only consider one learning material, lecture. We employ spectral clustering to help us evaluate the latent learning material features matrix obtained by our model. We assume that learning materials with similar features will be clustered into same group. 
% We observed that MORF\_QL dataset include learning materials from totally 8 weeks, then it is best to set the number of cluster equal to 8, since in most of scenario, learning materials from same week are in similar topics or they have relationship.
Figure~\ref{fig:clusters without quiz our} shows the 8 clusters that were discovered using MVKM, and Figure~\ref{fig:clusters without quiz LDA} shows the 8 clusters that were discovered using video-lecture transcripts.
Each cluster is shown within a box with a number associated with it.
Each video-lecture is shown by its module (or week in the course), its order in the module sequence, and its name.
For ease of comparison, we colored the video names according to their LDA content clusters.
Looking at the LDA content clusters, we can see that although some lectures in same module fit in same cluster (e.g., videos 1, 2, 3, and 4 from week 7 are all in cluster 7), some of the lectures do not cluster with other videos in their module.
For example, video 5 in week 7 is in cluster 2, with pioneer knowledge tracing methods. 
This shows that in addition to structural similarities, content similarities also exist in learning materials.
Looking at MVKM clusters, we can see that the clusters mostly represent the course structure similarity: learning materials from same module are grouped. 
%MORF\_QL dataset are collected from 8 weeks, therefore, we set the number of cluster be 8 to train clusters. 
%The clustering results are showed in figure \ref{fig:clusters without quiz our}. As we can see, most of lectures are clustered in the same group if they are in sequential (lecture videos are in the order of time they are presented). 
For example, all videos of week 3 are grouped in cluster 2. 
% -cluster 3 contains all videos of weeks 5 and 6, and initial videos of week 7;
% and cluster 8 includes the last videos of week 7 and videos of week 8.
However, we can see that in many cases, whenever the structure similarity in clusters are disrupted, it is because of the content similarity in video lectures.
% -However, we can see that in many cases, the content similarity is also captured in clusters.
% Particularly, whenever the structure similarity in clusters are disrupted, it is because of the content similarity in video lectures.
For example, video 5 in week 7 that was clustered with pioneer knowledge tracing method in LDA content clusters is also clustered with them in MVKM clusters.

\noindent\textbf{Between-Type Concept Evaluation. }To evaluate MV-KM's discovered similarities between different types of learning materials, we evaluate assignments' and video lectures' in MORF\_QL.
To do this, we build LDA-based clusters using assignment texts and video lecture transcripts.
These clusters are shown in Figure~\ref{fig:clusters with quiz LDA}.
We also cluster the learning materials using spectral clustering on the concatenation of their $Q^{[r]}$ matrices (Figure~\ref{fig:clusters with quiz our}).
Because the assignments bring more information to the clustering algorithms, the clustering results are different from the clusters of video lectures only.
%Next, we use learning material features and topic distributions of lectures as well as assignments to obtain clusters with spectral clustering method.
% we aggregate assignments and lectures together and using spectral clustering train clusters both by using the learning material feature and topic distribution from LDA again. 
%Figure \ref{fig:clusters with quizzes} exhibition these two clustering results.
Similar to within-type concept evaluation results, we can still see the effect of both content and structure similarities in video lectures that are clustered together by MVKM. 
For example, videos 1 and 3 of week 2 are clustered with later weeks' videos because of content similarity (cluster 1 in Figure~\ref{fig:clusters with quiz our}).
% -While videos 2 of week 2 is also clustered with them because of structural similarity, as it comes between these two videos in course sequence.
While videos 2 of week 2 is also clustered with them because it comes between these two videos in course sequence.

Additionally, between video lectures and assignments, the clusters closely follow the course structure.
The assignments in this course come at the end of their module and right before the next module starts.
For example, ``Assignment 3'' appears after video 5 at week 3 and before video 1 at week 4.
We can see that all assignments, except ``Assignment 1'' that is the first one, are clustered with their immediate next video lecture.
% -``Assignment 1'' has the most attempts, compared to the other video lectures and assignments in this course.
% -It is also not very similar to the other learning materials in terms of its content.
% -This can explain why it has not been clustered with week 2 video lectures.
%After adding the assignments, both clustering results are altered. and the reason is assignment will make the sequence of learning materials be different (between two lecture videos, there may be an quiz exist), and bring more content information than only considering lectures. 
%Even though, we still can observe that by using learning material feature of proposed model, lecture videos and assignment are roughly clustered according to the sequence that they are presented. 
Moreover, we can see the effect of content similarity between assignments and video lectures in differences of Figures~\ref{fig:clusters without quiz our} and ~\ref{fig:clusters with quiz our}.
For example, without including assignments, ``Week 1 Introduction'' and ``W1 V1: Big Data in Education'' were clustered together in cluster 7 of Figure~\ref{fig:clusters without quiz our}.
However, after adding assignments, because of the content similarity between ``Assignment 3'' and ``Week 1 Introduction'' ( Figure~\ref{fig:clusters with quiz LDA} cluster 2), ``Week 1 Introduction'' and ``W1 V1: Big Data in Education'' are clustered with video lectures that are structurally close to ``Assignment 3''.
% -In other words, they are clustered with videos of week 4 and last video of week 3 in cluster 2 of Figure~\ref{fig:clusters with quizzes} (left).

%For those lecture videos are not clustered into a cluster that they has sequential correlation with other group members, it is also efficient to  interpret this by the same reason that in the cluster they belong to, there are some lecture videos/assignments share similar topics with them. 
Altogether, we demonstrated that learning materials' bias parameters in MVKM are aligned with their difficulties; learning materials' latent concepts discovered by our model well represent learning materials' real-world similarities, both in structure and in content; and 
MVKM can successfully unveil these similarities between different types of learning materials, without observing their content or structure.

% \begin{figure}[h]
% \centering
% \includegraphics[width = .45\textwidth]{pics/wordcloud.pdf}
% \caption{Wordcloud of keyword for eight topics trained by LDA}
% \label{fig:wordcloud}
% \end{figure}

\section{Conclusions}
\vspace{-3pt}
\label{sec:conclusions}
In this paper, we proposed a novel Multi-View Knowledge Model (MVKM) that can model students' knowledge gain from different learning materials types, while simultaneously discovering materials' latent concepts.
Our proposed tensor factorization model explicitly represents students' knowledge growth and allows for occasional forgetting of learned concepts. Our extensive evaluations on synthetic and real-world datasets show that MVKM outperforms other baselines in the task of student performance prediction, can effectively capture students' knowledge growth, and represent similarities between different learning materials types.

\section{Acknowledgments}
This paper is based upon work supported by the National Science Foundation under Grant No. 1755910.
%\subsection{References}
% \clearpage
\bibliographystyle{abbrv}
\bibliography{sigproc}

\end{document}